\documentclass[12pt]{article}
\usepackage{a4wide}
\usepackage{amssymb,amsmath}

\usepackage{pictexwd,dcpic}

\newcommand{\fund}[2]{\ensuremath{\frac{\delta #1}{\delta #2}}}
\newcommand{\pard}[2]{\ensuremath{\frac{\partial #1}{\partial #2}}}
\newcommand{\gul}[2]{\ensuremath{H^{#1}_{#2}}}

\newcommand{\vd}[2]{\frac{\delta #1}{\delta #2}}   
\newcommand{\Par}[2]{\frac{\partial #1}{\partial #2}}   
\newcommand{\CPar}[2]{\frac{{\rm D} #1}{{\rm D} #2}}   

\newcommand{\Ef}{E^\varphi}
\newcommand{\Kf}{K_\varphi}
\newcommand{\md}{{\mathrm d}}

\newcommand{\li}{ [ \hspace{-0.7ex} [ \hspace{0.5ex} }     
\newcommand{\ri}{ \hspace{0.5ex} ] \hspace{-0.7ex} ] \hspace{0.5ex} }

\begin{document}
{\renewcommand{\thefootnote}{\fnsymbol{footnote}}
\medskip
\begin{center}
{\LARGE Discreteness corrections and higher spatial derivatives\\[3mm] in effective canonical quantum gravity}\\
\vspace{1.5em}
Martin Bojowald,$^1$\footnote{e-mail address: {\tt bojowald@gravity.psu.edu}}
George M. Paily$^1$\footnote{e-mail address: {\tt
    gpaily@allegheny.edu}. Present address: Allegheny College, Meadville, 
  PA 16335, USA}
and Juan D. Reyes$^2$\footnote{e-mail address: {\tt jdrp75@gmail.com}. Present
  address: Centro de Ciencias Matem\'aticas, 
Unidad Morelia, Universidad Nacional Aut\'onoma de
M\'exico, UNAM-Campus Morelia, A. Postal 61-3, Morelia, Michoac\'an 58090,
Mexico. 
Instituto de F\'{\i}sica y
Matem\'aticas,  Universidad Michoacana de San Nicol\'as de
Hidalgo, Morelia, Michoac\'an, Mexico.}  
\\
\vspace{0.5em}
$^1$Institute for Gravitation and the Cosmos,\\
The Pennsylvania State
University,\\
104 Davey Lab, University Park, PA 16802, USA\\
\vspace{0.5em}
$^2$Departamento de F\'isica, \\
Universidad Aut\'onoma Metropolitana-Iztapalapa, \\ 
San Rafael Atlixco 186, M\'exico D.F. 09340, M\'exico.\\
\vspace{1.5em}
\end{center}
}

\setcounter{footnote}{0}

\begin{abstract}
  Canonical quantum theories with discrete space may imply interesting
  effects. This article presents a general effective description, paying due
  attention to the role of higher spatial derivatives in a local expansion and
  differences to higher time derivatives. In a concrete set of models, it is
  shown that spatial derivatives one order higher than the classical one are
  strongly restricted in spherically symmetric effective loop quantum
  gravity. Moreover, radial holonomy corrections cannot be anomaly-free to
  this order.
\end{abstract}

\section{Introduction}

A canonical quantization of gravity implies different types of modifications
of the classical space-time continuum, depending on which precise methods are
used. In several approaches, discrete structures appear which should modify
not only the dynamics of the theory but also its fundamental symmetries. The
consistency of such modifications largely remains to be explored, especially
regarding conditions for a quantum theory free of gauge anomalies. Several
models of consistent \cite{UniformDisc} or perfect discretizations
\cite{PerfectAction,BrokenAction} are available. At the level of effective
actions or constraints, however, only pointwise quantum-geometry modifications
have been implemented so far, which may indirectly be related to spatial
discreteness but remain local and do not give rise to a derivative
expansion. In the case of cosmological perturbations in effective models of
loop quantum gravity, for instance, current versions either use pointwise
exponentials of a connection instead of integrated holonomies
\cite{VectorHol,ScalarHol}, or --- so far for vector modes only --- perform an
expansion but truncate it at one order before higher spatial derivatives would
become relevant \cite{VectorSecond}. The form or even the possibility of
consistent versions of discreteness in effective theories of loop quantum
gravity therefore remains unclear.

Motivated by this crucial gap in the current understanding of canonical
quantum gravity, we start in this article a systematic investigation of
consistent discreteness corrections using effective methods. These new
canonical tools are general, but we aim to provide specific examples in models
of loop quantum gravity, the canonical theory in which the most details about
discrete representations are available. In particular, in this setting there
are not only standard discretization effects of spatial derivatives, but also
a new type of modification related to the prominence of holonomies in its
kinematical quantum representation.

Loop quantum gravity \cite{Rov,ThomasRev,ALRev} is based on a representation
in which holonomies of connections rather than connection components act as
operators. This feature, as often emphasized is crucial for the spatial
background independence realized in the theory, and for the implication of
discrete spatial quantum geometry. Classical expressions that depend on
connection components, especially the Hamiltonian constraint of canonical
gravity, must then be regularized or modified before they can be turned into
operators, reflecting more-indirect discreteness effects in space-time
observables. Characteristic corrections to the classical dynamics result,
which have been investigated in quite some detail in loop quantum cosmology
and also in black-hole models. Most of these investigations, however, have
made use either of a complete elimination of local degrees of freedom in
exactly homogeneous minisuperspace models, or of gauge fixings or other
limitations on space-time structures such as specific choices of time
variables (deparameterization). In such models, it is not clear whether the
quantum theory analyzed is free of anomalies and whether it respects
covariance (or even energy conservation \cite{Energy}). That crucial space-time
effects may be missed by such restricted treatments is shown by a recent
discovery suggesting that space-time turns Euclidean when holonomy effects are
significant \cite{Action}. It is therefore incorrect to assume classical
properties of space-time, as one does when one fixes the gauge before
quantization or uses other properties of the classical constraints.

In order to overcome these limitations, one must find a consistent quantum
realization of the constraint algebra that remains first class off-shell. For
holonomy corrections, such systems have been found only in $2+1$ dimensions
\cite{ThreeDeform} or partially, and with effective methods, in spherically
symmetric models \cite{JR,LTBII} and for cosmological perturbations
\cite{ScalarHol}. Effective methods, developed for this purpose in
\cite{ConstraintAlgebra,ScalarGaugeInv}, allow one to compute Poisson brackets
of quantum-corrected constraints instead of commutators of constraint
operators, implying large simplifications. But still, the status of consistent
holonomy-corrected constraints remains incomplete even in symmetric models: So
far, only ``pointwise'' holonomies have been implemented, exponentiating
connection components without including spatial integrations along curves. The
weak non-locality associated with curve integrations turns out to be difficult
to parameterize and implement, even if one approximates it by a derivative
expansion. However, it is the most characteristic consequence of spatial
discreteness in loop quantum gravity. This non-locality, or higher spatial
derivatives, must therefore be realized in an anomaly-free way before one can
be sure that discrete quantum space as envisaged by loop quantum gravity is
able to provide consistent space-time models.

In this article, we investigate these issues further. We first provide a
systematic treatment of a derivative expansion in effective canonical gravity,
taking into account the different appearances of time and space derivatives in
a Hamiltonian setting. To facilitate such expansions for first-class
constraints, subject to the strong requirement of anomaly-freedom of any
correction terms, we will provide several formulas of Poisson brackets
applicable generally. Our main examples will be given in spherically symmetric
models. So far the results are negative: in the models considered, higher
spatial derivatives one order above the classical one are ruled out. But given
the complexity of the problem, our analysis of spherically symmetric models of
loop quantum gravity remains incomplete, and there is still room for
potentially consistent versions based on more general (and more complicated)
parameterizations of discreteness effects.

\section{Holonomy corrections}

Discreteness corrections appear in different forms whenever a Hamiltonian is
to be formulated on a discrete structure. In general, the only feature one may
assume for an effective formulation is the presence of higher spatial
derivatives, resulting from finite differences expanded for a local effective
description. But there are also more characteristic consequences of some
approaches, most importantly the use of holonomies in loop quantum
gravity. Properties of holonomy operators and their conjugate fluxes imply
spatial discreteness at a kinematical level \cite{AreaVol}, and they lead to
modifications of the Hamiltonian constraint \cite{RS:Ham,QSDI} that have
consequences for the dynamics as well as the structure of space-time.

Holonomies $h_e(A_a^i)={\cal P}\exp(\smallint \tau_iA_a^i\dot{e}^a{\rm
  d}\lambda)$ (with Pauli matrices $2i\tau_j$) represent the su(2)-connection
$A_a^i$ by SU(2)-elements associated to all (analytic) curves $e$ in space. In
loop quantum gravity \cite{Rov,ThomasRev,ALRev}, they act as multiplication
operators which, starting with a connection-independent state, construct the
whole kinematical Hilbert space. Collecting all curves used in repeated
actions of holonomies, one obtains a graph as a model for the underlying
discrete space. Spin-network states can be used as a basis of the Hilbert
space \cite{RS:Spinnet}.

On these graphs, holonomies act by generating new edges or by changing the
excitation level (the spins) of existing ones. A holonomy around a closed loop
can be used to approximate curvature components of $A_a^i$, the better the
smaller the loop is. However, the limit of the holonomy around a loop
shrinking to a point divided by the coordinate area of the loop, in which case
one would obtain exactly the curvature components, does not exist for
operators: A holonomy operator maps a state into an orthogonal one, so that
the limit state in the attempted construction would be orthogonal to all
states in the limiting sequence, a contradiction. One can work only with
extended loops, so that classical expressions containing the connection or its
curvature must be modified before quantization by reexpressing them in terms
of extended holonomies. For low curvature and small loops, one would still be
close to the classical expressions, so that the semiclassical limit would not
be in danger.

One can develop a more refined argument if one exploits spatial diffeomorphism
invariance of the theory \cite{QSDI}. If one implements the diffeomorphism
gauge freedom before one constructs curvature-dependent operators such as the
Hamiltonian constraint, a small loop is gauge equivalent to a larger one. The
diffeomorphism-invariant state with a loop added does not change as the loop
is shrunk, and one can take the limit in a trivial way, the loop remaining
attached. Applying this construction to the Hamiltonian constraint, one can
argue that the algebra of constraint operators is anomaly-free on the space of
states solving the diffeomorphism constraint \cite{AnoFree}.

However, it is not clear what quantum corrections follow from these operators
because the diffeomorphism-invariant level makes it difficult to associate a
semiclassical or other geometry with these states. Moreover, these
constructions prevent one from addressing the full off-shell constraint
algebra, in which the Hamiltonian constraint is paired non-trivially with the
diffeomorphism constraint. Even though one ultimately wants to solve all
constraints and derive physical observables, the space of
diffeomorphism-invariant states is too restricted to analyze full consistency
of the theory or to uncover non-observable properties of conceptual interest,
such as the form of quantum space-time structure realized. To address these
questions, one must look for a consistent realization of all first-class
constraints relevant for the space-time gauge without solving any of
them. (Canonical formulations that make use of triad variables are subject to
a Gauss constraint. This constraint is not relevant for space-time structure
and can therefore be treated separately from the diffeomorphism and
Hamiltonian constraints. Moreover, it is simple enough to allow explicit
solutions and direct implementations in the quantum theory, without giving
rise to quantum corrections or deformations of its gauge structure.)

When both the Hamiltonian constraint and the diffeomorphism constraint remain
unsolved, looking for an off-shell realization of a first-class constraint
algebra becomes very difficult. One must then deal with the issue of
``structure functions'' in the bracket $\{H[N],H[M]\}=
D[h^{ab}(N\partial_bM-M\partial_bN)]$ of two Hamiltonian constraints, where
$h^{ab}$ is the inverse spatial metric. Moreover, the argument by which one
can eliminate the loop regulator exploiting diffeomorphism invariance no
longer applies. Holonomies are not just regulated versions of the connection;
they imply modifications of the classical dynamics which may be small at low
curvature but can still have important implications. If they break the closure
of the constraint algebra, the gauge structure will be violated no matter how
small possible modifications are. The theory will be anomalous and
inconsistent.

It is not clear at present whether a consistent off-shell realization of loop
quantum gravity exists. Models in which this has been achieved, so far in
$2+1$ dimensions at the operator level
\cite{ThreeDeform,TwoPlusOneDef,TwoPlusOneDef2,AnoFreeWeak}, show that a
consistent first-class algebra can be realized only if it is deformed, with
quantum corrections in its structure functions. These results, including the
specific form of corrections, are consistent with effective calculations, in
which one inserts possible quantum corrections in the classical constraints
and computes their Poisson brackets \cite{ConstraintAlgebra,JR,ScalarHol}. We
will make use of effective constraints in spherically symmetric models, which
strike a nice balance between interesting off-shell properties and rather
manageable calculations.

\subsection{Spherical symmetry}

A spherically symmetric SU(2) connection has the form
\cite{SymmRed,SphSymm,SphSymmHam,Springer}
\begin{equation}
 A_a^i\tau_i{\rm d}x^a= A_x\tau_3{\rm d}x+ A_{\varphi}\Lambda^A{\rm d}\vartheta+
 A_{\varphi}\bar{\Lambda}^A{\rm d}\varphi+\cos\vartheta\:\tau_3{\rm d}\varphi
\end{equation}
with an internal frame such that ${\rm tr}(\Lambda^A\tau_3)= {\rm
  tr}(\bar{\Lambda}^A\tau_3)= {\rm tr}(\Lambda^A\bar{\Lambda}^A)=0$. The
requirement that internal gauge transformations must fix $\tau_3$ reduces the
internal gauge group to U(1). The gauge is partially fixed at this stage, but
the Gauss constraint is so simple that one can easily demonstrate the
independence of quantum results of the chosen gauge fixing. Moreover, the
constraint is not modified by holonomy corrections: Its quantization makes use
only of invariant vector fields on spaces of connections. No anomalies are
introduced at this stage. Finally, the part of the gauge freedom fixed here
does not refer to space-time structure, and is therefore not crucial in the
present context.

A spherically symmetric densitized triad has the dual form
\begin{equation}
  E^a_i\tau^i\frac{\partial}{\partial x^a}=
  E^x\tau_3\sin\vartheta\frac{\partial}{\partial x}+
  E^{\varphi}\Lambda_E\sin\vartheta \frac{\partial}{\partial\vartheta}+
  E^{\varphi}\bar{\Lambda}_E\frac{\partial}{\partial\varphi}\,.
\end{equation}
Its coefficients determine the spatial metric
\begin{equation}
 {\rm d}s^2= \frac{(E^{\varphi})^2}{|E^x|} {\rm d}x^2+ |E^x| ({\rm
   d}\vartheta^2+\sin(\vartheta)^2 {\rm d}\varphi^2)\,.
\end{equation}
The internal triad $(\tau_3,\Lambda^E,\bar{\Lambda}^E)$ is independent of the
one in the connection, except that the same $\tau_3$ is used in both
cases. There is a free angle (denoted by $\beta$ in \cite{SphSymm,SphSymmHam})
to rotate the internal triads into each other, which together with its
momentum is an invariant kinematical degree of freedom, in addition to
$(A_x,E^x)$ and $(A_{\varphi},E^{\varphi})$, but is eliminated when the
remaining U(1) Gauss constraint is solved. While $A_x$ is canonically
conjugate to $E^x$, $E^{\varphi}$ is not conjugate to $A_{\varphi}$ but to the
extrinsic-curvature component $K_{\varphi}=-2\gamma^{-1}A_{\varphi} {\rm
  tr}(\Lambda^A\Lambda_E)$. The $x$-component of extrinsic curvature is
$K_x=\gamma^{-1}(A_x+\eta')$ with $\eta=-2{\rm tr}(\tau_3\Lambda_E)$. One can
obtain $K_x$ as a U(1)-gauge invariant combination of $A_x$ and $\eta$.  In
spherical symmetry, one can therefore easily work with extrinsic curvature
instead of connections, also in holonomies.

In addition to the appearance of U(1) instead of SU(2), a further
simplification of spherical symmetry is the form of graphs. Inhomogeneity is
realized only in the radial $x$-direction, along which one can align vertices
connected by links. Any extended holonomy simply integrates over a piece of
the $x$-axis, as in $\bar{h}_e(A_x)=\exp(\tau_3\smallint A_x {\rm d}x)$, or
$h_e(A_x)= \exp(i\smallint A_x{\rm d}x)$ for matrix elements exhibiting the
U(1) nature. (No path ordering is necessary for the Abelian reduced theory.)
In any U(1)-gauge invariant state, an $A_x$-holonomy is combined with point
holonomies for $\eta$ so that the state depends only on $A_x+\eta'$: It
suffices to look at the basis of charge-network states, in which each edge $e$
carries an integer charge quantum number $k_e$ for its $A_x$-holonomy, and
each vertex an integer quantum number $k_v$ for a point holonomy
$\exp(i\eta(v))$ if the U(1)-field $\eta$ (in addition to a real quantum
number $\mu_v$ for $K_{\varphi}(v)$, taking values in the Bohr
compactification of the real line).  The conservation of U(1)-flux at a vertex
$v$ with $\eta$-charge $k_v$ implies the relation $k_{e_+}-k_{e_-}+k_v=0$,
where $e_{\pm}$ are the two edges touching the vertex $v$ with charge labels
$k_{e_{\pm}}$. The corresponding (point) holonomies then appear in a
spin-network state as factors
\[
\cdots e^{ik_{e_-} \smallint^{v}A_x {\rm d}x} e^{ik_v\eta(v)} e^{ik_{e_+}
\smallint_{v}A_x {\rm d}x}\cdots= \cdots e^{ik_{e_-}
\smallint^{v}(A_x+\eta') {\rm d}x} e^{ik_{e_+} \smallint_{v}(A_x+\eta') {\rm
  d}x}\cdots\,.
\]
Gauge-invariant states therefore depend only on the combination
$A_x+\eta'=\gamma K_x$ of $A_x$ and $\eta$.  The other connection component
$A_{\varphi}$ or rather the $E^{\varphi}$-momentum $K_{\varphi}$ appears in
holonomies along curves in the $\varphi$-direction, along which $K_{\varphi}$
does not change. Such a holonomy is simply an exponential
$h_{(v,\delta)}(K_{\varphi})= \exp(i\delta K_{\varphi}(v))$ with $K_{\varphi}$
evaluated at a point (or vertex) $v$, and with a real number $\delta$ (or
possibly a function on phase space) related to the coordinate length of the
curve one would integrate over.

With these variables, one can check what kind of holonomy modifications are
possible in the Hamiltonian constraint so that a first-class algebra
results. The possibility of pointwise modifications in $K_{\varphi}$ has been
clarified \cite{JR}, with the result that they can leave the algebra
first-class but always deform it. The form of the deformed algebra is very
characteristic:
\begin{equation} \label{deformedSSalgebra}
\{H[N],D[N^x]\}=-H[N^xN'] \quad \text{ and }  \quad
\{H[M],H[N]\}=D[\beta|E^x|(\Ef)^{-2}(MN'-M'N)]\, 
\end{equation}
with $N^x$ the only non-vanishing component in the radial direction of the
shift vector, and $\beta$ a correction or deformation function depending on
phase space variables. This form also agrees with the one found in
$2+1$-dimensional models and for cosmological perturbations. However, no
consistent holonomy modification of the $K_x$-dependence has yet been
found. These corrections are more difficult to realize, not the least because
they require integrations, and therefore lead to either non-locality or
higher-derivative theories.

\subsection{Parameterization}

A single gauge-invariant combination of holonomies in spherical symmetry is
given by $h_e=\exp(i \gamma \smallint_{x_0}^{x_0+\ell_0} K_x {\rm d}x)$, where
$x_0$ is the starting point of the curve and $\ell_0$ its coordinate
length. As an operator, $\hat{h}_e$ will add an edge from $x_0$ to
$x_0+\ell_0$ to the graph underlying a state it acts on, or increase the
quantum numbers on pieces of a spin network overlapping with the curve from
$x_0$ to $x_0+\ell_0$. Composite operators depending on the connection, such
as the Hamiltonian constraint, make use of these basic holonomy operators, but
they usually come with a specific description of how the curves for holonomies
are chosen with respect to a graph state acted on. They may leave the graph
unchanged, making use only of holonomies along curves between vertices already
present in the original graph, or they may be graph-changing and create new
vertices. In the former case, $\ell_0$ for an individual holonomy in a vertex
contribution of the operator would be fixed as the coordinate distance to the
next vertex.

In the latter case, which is more complicated, but preferred in the full
theory for the arguments of anomaly-freedom on diffeomorphism-invariant states
to work, $\ell_0$ would not be constant; one would have to find an alternative
way to determine its values. One could, for instance, assume that the
graph-changing nature of the operator leads to dynamical lattice refinement so
that the geometrical length 
\begin{equation}\label{ell}
 \ell=\int_{x_0}^{x_0+\ell_0}\sqrt{g_{xx}}{\rm d}x\approx
 \ell_0\sqrt{g_{xx}}=\ell_0  E^{\varphi}/\sqrt{|E^x|}\,,
\end{equation}
measured with the densitized triad or the metric component $\sqrt{g_{xx}}=
E^{\varphi}/\sqrt{|E^x|}$, has a certain dependence on geometrical variables
such as the orbit area $|E^x|$. (We assume $\ell_0$ to be sufficiently small
compared to the scale on which $g_{xx}$ varies. If this assumption is violated
or not precise enough, a derivative expansion of the integral can be used, as
described in more detail below.)  The simplest possibility in this context
would be for $\ell$ (rather than $\ell_0$) to be some constant, such as the
Planck length, but this is not the only choice.

A constant $\ell$ would be analogous to a certain class of cosmological models
\cite{APSII} often studied in loop quantum cosmology, which is also shown by
the behavior of holonomies. To see this, we assume that we are close to
homogeneous models, so that $\ell_0$ for a given holonomy may be very short
compared to $K_x/K_x'$, which is generically large for $K_x'$ restricted to be
small by near homogeneity.  The dominant contribution to the argument
$\int_{x_0}^{x_0+\ell_0}K_x{\rm d}x$ of the exponential in a holonomy is then
simply
\begin{equation}
 \ell_0K_x(x_0)\approx \ell \frac{\sqrt{|E^x|}}{E^{\varphi}} K_x\,.
\end{equation}
According to the classical equations of motion, which may be used when the
present assumption is satisfied and $K_x$ is small\footnote{Assuming that we
  are close to a homogeneous model restricts the possible choices of
  space-time slicings, so that a ``low-curvature'' regime may be demarcated in
  terms of the non-invariant curvature component $K_x$. Subtleties in the
  general case of inhomogeneous --- but still spherically symmetric ---
  geometries will be discussed below.} compared to $1/\ell_0$, we write
\begin{equation}
\ell \frac{\sqrt{|E^x|}}{E^{\varphi}} K_x = -\ell
\left(\frac{\dot{E}^{\varphi}}{E^{\varphi}}-
  \frac{1}{2}\frac{\dot{E}^x}{E^x}\right)
= -\ell
\frac{(E^{\varphi}/\sqrt{|E^x|})^{\bullet}}{E^{\varphi}/\sqrt{|E^x|}}= -\ell
\frac{\sqrt{g_{xx}}^{\bullet}}{\sqrt{g_{xx}}}\,.
\end{equation}
For a cosmological model, $\sqrt{g_{xx}}=a$ would be the scale factor, so that
the argument of holonomies agrees with $\ell {\cal H}$, using the Hubble
parameter ${\cal H}=\dot{a}/a$.

Different parameterizations (or lattice refinement schemes
\cite{InhomLattice,CosConst}) are possible in which $\ell$ is not constant
but, for instance, a certain power of $|E^x|$ or of $g_{xx}$, or some other
function. We will not assume any specific function but simply take into
account the fact that the choice of routings of curves may lead to a triad
dependence of holonomies in addition to the expected connection or
extrinsic-curvature dependence.

\subsection{Derivative expansion}
\label{s:Deriv}

In the previous subsection, we assumed, restricting ourselves to
near-homogeneous low-curvature geometries, that the coordinate length $\ell_0$
is sufficiently small, so that holonomy corrections would be weak. As we
approach regimes in which quantum geometry is more pronounced, stronger
modifications of the dynamics arise from the use of holonomies and
higher-order corrections must be taken into account. In an inhomogeneous
model, not just higher powers of $\ell_0\gamma K_x$ in an expansion of the
pointwise
\[
 \exp(i\ell_0\gamma K_x)=\sum_{n=0}^{\infty}\frac{1}{n!} (i\ell_0\gamma K_x)^n
\]
will grow, but also higher spatial derivatives of $K_x$ in a derivative
expansion of the integrated $\int_{x_0}^{x_0+\ell_0} K_x{\rm d}x$. 

The treatment of expansions now becomes more subtle, related to the
interpretation of the non-invariant $K_x$ as some kind of measure for
curvature, or at least as a parameter that tells us when holonomy corrections
become large. The extrinsic-curvature component $K_x$ depends on the slicing,
and one may locally be able to make it small in high-curvature regimes, or to
make it large even in flat space-time, just by choosing an appropriate slicing
of space-time. It therefore seems inconsistent to use its magnitude to
determine the strength of corrections or orders of expansions. This problem,
of course, does not arise just in effective descriptions; it plays a role
already in constructing a full Hamiltonian constraint \cite{RS:Ham,QSDI}, in
which one replaces connections by holonomies so that the classical expression
is obtained in the ``classical limit,'' or for ``small'' connections. (In the
complete classical limit, loops shrink to points and connections can take
arbitrary values. But for correct semiclassical physics, corrections to the
classical limit must be small, which can be realized only for sufficiently
small connections.) The problem we encounter in formulating a systematic
derivative expansion is therefore a more general one: It arises because
canonical quantum gravity primarily implies modifications for the Hamiltonian,
rather than for space-time covariant expressions such as an action with its
coordinate-independent meaning.

On closer inspection, it turns out that the form of discreteness corrections
expressed in terms of extrinsic curvature is well-defined. A key role in this
argument is played by the fact that, as shown in (\ref{deformedSSalgebra}),
holonomy modifications cause corrections of the algebra of hypersurface
deformations by an additional function $\beta$ depending on extrinsic
curvature. The algebra of constraints still closes, and therefore deviations
from the classical value $\beta=1$ have invariant meaning. This statement is
clear from the algebra, but it seems surprising given the non-invariant form
of extrinsic curvature. Several further implications of modified constraints
then come into play: First, modified constraints imply corrections in the
classical equations of motion, and canonically, the role of K-components as
extrinsic curvature follows only after the Hamiltonian equation of motion for
the triad is used. With modified equations, $K$-components no longer are
extrinsic curvature in the classical sense, and classical intuition about the
values $K$-components may take for different slicings breaks down. Secondly, a
deformed algebra means that gauge transformations, although not violated, no
longer generate space-time Lie derivatives or changes of slicings in classical
space-time. Therefore, what one may achieve for $K$-components in classical
space-time plays no role for possible $K$-values in a deformed system. Two
slices of the same classical space-time, one with large $K$ and one with small
$K$, do not produce gauge-related solutions of the modified system. The fact
that one solution would deviate more strongly from the classical space-time
than the other one is not a contradiction. These arguments further highlight
the importance of deformed algebras and their derivation (but also the dangers
of using too much classical space-time intuition when one interprets canonical
quantum gravity). They allow us to quantify the strength of holonomy
corrections in terms of ``extrinsic-curvature'' components, and to organize
expansions.

In general, then, one expects that holonomy corrections become strong at high
curvature. As one leaves the classical regime, deviations from both the
dynamics and the form of space-time will grow. While classical intuition will
break down at some point before the Planck regime is reached, effective
equations allow one to study the consequences of quantum aspects. For leading
orders of an expansion, one may still use classical expectations to estimate
what correction terms are relevant, and these should be all terms that
contribute to curvature invariance of some order.  Relevant terms may equally
result from $K_x^n$ (a higher power) or $K_x^{(n)}$ (a higher derivative), or
combinations thereof. Riemann curvature is, after all, a sum of powers of the
space-time connection and its derivatives. Unless one works in a specific
space-time gauge or slicing, which usually is not legitimate before
quantization, one cannot assume that only powers $K_x^n$ but no higher spatial
derivatives $K_x^{(n)}$ should contribute. (Time derivatives play a special
role in a canonical theory. We will discuss them below.)

The classical constraint is therefore modified not only by pointwise holonomy
corrections of the form $\exp(i\ell_0\gamma K_x(x_0))$ but also by higher
spatial derivatives of $K_x$. For any explicit local effective constraint, a
combined expansion is required, one of the form
\begin{eqnarray}
  \exp\left(i\gamma \int_{x_0}^{x_0+\ell_0} K_x{\rm d}x\right) &=&
  \exp\left(i\gamma \int_{0}^{\ell_0}(K_x(x_0)+ h
  K_x'(x_0)+{\textstyle\frac{1}{2}}h^2K_x''(x_0)+\cdots){\rm d}h\right) 
 \nonumber\\
  &=& \exp\left(i\gamma (\ell_0K_x(x_0)+ \frac{1}{2}\ell_0^2 K_x'(x_0)+
  \frac{1}{6} \ell_0^3 K_x''(x_0)+\cdots)\right)\nonumber\\
  &=& 1+i\ell_0\gamma K_x(x_0) + \frac{1}{2}\ell_0^2(i \gamma K_x'(x_0)- 
  \gamma^2K_x(x_0)^2)\nonumber\\
 &&+
  \frac{1}{6} \ell_0^3 (i\gamma K_x''(x_0)-3\gamma^2 K_x(x_0)K_x'(x_0)-
i\gamma^3K_x(x_0)^3
  )+\cdots\,.\label{Kderiv}
\end{eqnarray}
(Note that this derivative expansion, unlike the continuum limit of the
difference equation for states \cite{cosmoIV,SemiClass} in homogeneous models,
is not controlled by $\gamma$.)  The expansion by powers of $\ell_0$ ensures
that all terms with the same number of derivatives are grouped together
provided we count the time derivative implicit in $K_x$ once equations of
motion are used.  

If $\ell_0$ is not fixed but triad-dependent as per (\ref{ell}), it may be
evaluated at $x_0$ or at $x_0+\ell_0$ or some other point, so that a further
derivative expansion of $\ell_0$ and the triad variables it may contain would
have to be included. (A derivative expansion also contributes higher-order
terms to the integral in (\ref{ell}).) The specific evaluation point depends
on how lattice refinement implies a triad-dependent $\ell_0$. Parameterizing
the evaluation point as $x_0+r\ell_0$ with some $0\leq r\leq 1$, we can write
\begin{equation} \label{Ederiv}
 \ell_0(E(x_0+r\ell_0))= \ell_0(E(x_0))+ r\ell_0 ({\rm d}\ell_0/{\rm
   d}E)|_{E(x_0)} E'(x_0) +\cdots\,.
\end{equation}
If $\ell_0$ is small and lattice refinement (or whatever causes the triad
dependence of $\ell_0(E)$) is not too violent, ${\rm d}\ell_0/{\rm d}E$ is
small as well, so that the derivative term in (\ref{Ederiv}) is of second
order in our expansion. For instance, for a power law $\ell_0(E)\propto E^y$,
as realized in all cases studied so far, including \cite{APSII}, we have ${\rm
  d}\ell_0/{\rm d}E\propto y\ell_0/E$, and we can write the second-order term
as $ry\ell_0^2E'/E$. Lattice refinement or the evaluation point of $\ell_0$
therefore determine the coefficients of the derivative expansion, but do not
affect its general form.  Here, for the general consideration of effective
descriptions, it is sufficient to know that we should expect the appearance of
further triad derivatives in an expansion whose coefficients, such as $r{\rm
  d}\ell_0 /{\rm d}E$ in (\ref{Ederiv}), so far remain undetermined by a
derivation from the full theory. The purpose of our constructions is to derive
possible restrictions on such coefficients using only the requirement of
anomaly-freedom.

\subsection{Holonomy corrections vs.\ higher-curvature corrections}

Holonomy corrections, according to (\ref{Kderiv}), imply higher powers of the
connection or extrinsic curvature in the Hamiltonian constraint, as well as
higher spatial derivatives. These features are shared with higher-curvature
corrections in an effective action, except that higher time derivatives always
come along with higher curvature but are not suggested by holonomy
corrections. Holonomy corrections are indeed different from higher-curvature
ones; they result from a modification of the Hamiltonian constraint motivated
by {\em spatial} quantum geometry, not from generic {\em space-time} covariant
correction terms in the form of higher curvature invariants. Even though
holonomy corrections and higher-curvature corrections are expected to be
significant in the same regimes --- when curvature reaches Planckian values
--- they must be distinguished from each other both in their formal derivation
and in their possible implications.

Holonomy corrections and higher-curvature corrections have very different
effects on quantum space-time structure. Generic higher curvature corrections
are determined by all possible terms that could modify the classical action by
higher derivatives in a covariant way, leaving the classical space-time
structure unchanged. Only the dynamics is then modified at high
curvature. Generic holonomy corrections, on the other hand, are introduced at
the kinematical level in order to quantize the Hamiltonian constraint. The
constraints themselves determine what space-time structure is realized, by
generating gauge transformations that classically correspond to space-time Lie
derivatives of phase-space functions. When quantum corrections are inserted in
the constraints, their transformations change. Gauge transformations could be
violated, and in general will be unless one is very careful about arranging
different correction terms. If this happens, the theory is anomalous and
inconsistent because its equations do not have mutually compatible
solutions. (One would obtain formally consistent solutions if one solves the
constraint as second-class ones. But then there are neither second-order
equations of motion nor gauge transformations that would remove spurious
degrees of freedom.)  In the consistent case, when all quantized constraints
still generate gauge transformations and remain first class, their algebra,
not just their functional form, in general carries quantum corrections. Their
gauge transformations no longer correspond to Lie derivatives on the
constraint surface, which implies that the space-time structure is changed by
quantum corrections. This consequence is realized for holonomy modifications
in all consistent versions found so far. Unlike higher-curvature corrections,
they modify the notion of space-time and general covariance. At high density,
modifications can be so strong that space-time turns into a quantum version of
4-dimensional Euclidean space.

Regarding their formal derivation, holonomy corrections and higher-curvature
terms make use of different mathematical structures and expansions. They also
imply different changes of the number of local degrees of freedom.

\subsubsection{Derivatives} 

As already discussed, higher spatial derivatives result from holonomy
corrections if a derivative expansion is used to approximate their
integrations locally.  Spatial derivatives of the connection (or extrinsic
curvature) and, if curve lengths are taken as triad-dependent, of the triad
result. Since $E^a_i$ and $K_a^i$, at this stage, both are phase-space
functions, one could expect them to appear in integrations
$\int_{x_0}^{x_0+\ell_0(E)}K_x{\rm d}x$ on the same footing, organized in a
derivative expansion by increasing orders $n$ of derivatives $K_x^{(n)}$
grouped with $E^{(n)}$. After all, the implicit time derivative contained in
$K_x$ can be seen only when equations of motion are used, but the latter are
not available before the Hamiltonian constraint is quantized and imposed. They
cannot be used for an analysis of the off-shell constraint algebra necessary
to study possible anomalies.

Nevertheless, we have already seen in (\ref{Kderiv}) that the expansion is
arranged as if the time derivative implicit in $K_x$ were present, if we just
expand by powers of the edge length $\ell_0$. Similarly, if we expand a
triad-dependent $\ell_0$ as in (\ref{Ederiv}), every factor of $\ell_0$ comes
along with an additional spatial derivative of $E^x$. This expansion therefore
treats $K_x$ and $(E^x)'$, or in general $K_x^{(n-1)}$ and $(E^x)^{(n)}$ on
the same footing. Even if we do not refer to implicit time derivatives or
equations of motion, all derivatives are ultimately counted.

For higher-curvature corrections, one considers as being on the same footing
all derivatives that would contribute to a curvature invariant of some
order. Since equations of motion would classically tell one that $K_x$ is
related to a first-order time derivative of the triad, a derivative expansion
is automatically organized by increasing orders $n$ of derivatives
$K_x^{(n-1)}$ and $E^{(n)}$, taking into account the extra time derivative
already present in $K_x$. Derivative expansions for holonomy corrections and
higher-curvature corrections therefore combine terms in the same manner.

Still, it is a priori unclear in which way one should organize the derivative
expansion of constraints, treating the $K$-components as derivatives or
not. The two different types of brackets in the hypersurface-deformation
algebra involving the Hamiltonian constraint $H[N]$, $\{H[N],D[N^x]\}$ and
$\{H[N],H[M]\}$, seem to require different viewpoints. The bracket
$\{H[N],D[N^x]\}$ being first class makes sure that $H[N]$ transforms
according to some consistent spatial geometry, which would be just the
classical one if we use an unmodified diffeomorphism constraint $D[N^x]$ (as
we will do below; see also \cite{Action}). A closed bracket then requires all
terms in $H[N]$, including its corrections, to combine to scalars of the
correct spatial density weight. For the latter, only spatial derivatives count
but not the time derivatives implicitly contained in the $K$-components. (Some
of the phase-space variables carry intrinsic density weights, with $K_x$ and
$E^{\varphi}$ of density weight one. From the viewpoint of spatial
diffeomorphisms, $K_x$ should therefore be on the same footing as
$E^{\varphi}$, not as $K_{\varphi}$ as the implicit time derivative would
suggest.) For $\{H[N],H[M]\}$, on the other hand, closure implies consistent
space-time dynamics, expected to be at least partially of higher-curvature
type. Here, it would seem more natural to count $K$-components as first-order
(time) derivatives.

We will for now avoid making a fixed choice on the order of derivatives and
their counting. It turns out that the specific form of variational methods in
this context, discussed further below, offers further insights and
guidelines. In particular, the order of derivatives of multipliers $N$ and
$N^x$, not just those of phase-space variables, plays an important role in
organizing expansions of Poisson brackets in the hypersurface-deformation
algebra.

\subsubsection{Degrees of freedom}

Holonomy corrections just modify the dependence of the Hamiltonian constraint
(or its expectation values used for effective constraints) on the
connection. No additional degrees of freedom are implied since only the
connection and the triad are quantized, providing the same number of basic
operators that exist as basic classical phase-space variables.

Higher-curvature corrections imply higher time derivatives and therefore new
degrees of freedom if initial values for higher-derivative equations are to be
imposed. Interpreting higher-derivative equations perturbatively, the number
of independent solutions does not change because extra solutions beyond the
classical number would be non-analytic in the perturbation parameter and must
be discarded for consistency \cite{Simon}. Nevertheless, higher time
derivatives imply corrections which can be understood as coupling terms with
these new, virtual degrees of freedom, just as virtual particles imply quantum
corrections in perturbative quantum field theory. Canonically, the new degrees
of freedom take on a much more explicit form \cite{EffAc,Karpacz}: they arise
as fluctuations and higher moments of a quantum state, parameters which are
independent of expectation values of basic operators to which the classical
phase-space structure can be applied. In certain regimes, these moments,
provided they change slowly, can be solved for in terms of expectation values
and inserted into expectation-value equations. In this way, the coupling terms
implicitly realized in higher-derivative equations become explicit
\cite{HigherTime}.

\subsubsection{Algebra}

In a complete semiclassical or effective expansion of a loop-quantized theory,
both holonomy corrections and higher time derivatives, resulting from
couplings to moments of a state, are present. Moreover, because of their
relation to curvature they are both expected to be significant in the same
regimes and cannot easily be separated from each other. Only a combined
treatment including both types of corrections can be fully consistent. Thanks
to their different formal and space-time roles, however, one can easily
separate these two modifications in formal derivations. 

Formally, holonomy corrections modify the dependence of constraints on
classical variables, while higher time derivatives come from moments of a
state. The gauge transformations they generate (if they indeed do generate
gauge) therefore affect different degrees of freedom. While a constraint
modified only by holonomy corrections implies modified gauge transformations
for expectation values, it leaves moments of canonical basic operators
invariant. A constraint modified by moments or higher time derivatives, on the
other hand, always generates gauge transformations that change the moments as
well. In this way, considering not just the magnitude of typical correction
terms but also the form of the modified gauge theory, one can keep holonomy
corrections and higher-curvature ones separate from each other.

Gauge transformations of the constraints of gravity encode the form of the
space-time structure realized. Since the transformations change in different
and distinguishable ways for the two types of curvature-related corrections,
taken separately they imply different space-time structures. Higher-curvature
terms, by definition, leave the classical space-time structure and the notion
of general covariance unchanged. Holonomy corrections, in all consistent
versions found so far, modify space-time structure and covariance. These
modifications, in general, cannot be canceled by higher time derivatives (or
other quantum-geometry corrections such as inverse-triad terms), and therefore
the space-time structure following from holonomy corrections alone is a good
indication of what a combined system would imply. If an anomaly-free version
of holonomy-modified constraints can be found, it will certainly provide a
consistent space-time model. For this reason, we focus on holonomy corrections
in this paper (but take along inverse-triad corrections), leaving out moment
terms which are more difficult to derive.

\section{Constraint algebra}

As indicated by the prevalence of deformed constraint algebras in loop quantum
gravity, we are in a situation much more general than the one of standard
higher-curvature effective actions. The latter, even though they may modify
the classical dynamics considerably, all have the same classical
hypersurface-deformation algebra for their constraints
\cite{HigherCurvHam}. Models of loop quantum gravity implement quantized
space-time structures, while higher-curvature effective actions take into
account modified dynamics of a standard space-time. This difference
has an influence on the derivation of possible consistent constraint algebras:
While higher-curvature actions always produce the classical bracket
$\{H[N],H[M]\}= D[h^{ab}(N\partial_bM-M\partial_aN)]$ with only first
derivatives of the multipliers, integrations by parts applied to some
$\{H[N],H[M]\}$ with constraints modified by higher spatial derivatives should
in general produce terms with as many derivatives of $N$ and $M$ as assumed in
a derivative expansion. Correspondingly, additional consistency conditions may
be obtained by requiring the algebra to close to all orders considered.

\subsection{General procedure}

In the presence of higher spatial derivatives, derivatives of the multipliers
$N$ and $N^x$ may be obtained in Poisson brackets, which raises the question
in how far multipliers and their derivatives can be treated as
independent. Using integrations by parts, a single constraint such as $H[N]=0$
can be rewritten in such a form, that derivatives of $N$ appear in the
integrand. Such mere rewritings, schematically $H[N]=H_1[N]+H_2[N']$, clearly
cannot lead to additional constraints because there was just one constraint to
begin with. Indeed, one cannot treat $N$ and $N'$ as independent and derive
two constraints $H_1=0$ and $H_2=0$ from the one original $H=0$: The local
constraints on phase-space functions are obtained by requiring $H[N]=0$ for
all {\em functions} $N$. The function itself and its derivatives (as opposed
to their values at a single point) are not independent, and therefore no
additional constraints arise by applying integrations by parts.

These circumstances are rather obvious and often used at least implicitly when
dealing with smeared constraints $H[N]$. One may employ them to reduce the
freedom in writing the constraints: If we require that only the multiplier $N$
but none of its spatial derivatives appear in the constraint expression, the
freedom of integrations by parts is strongly reduced. This condition could not
be used if quantum gravity or some other effects would give rise to
corrections with higher spatial derivatives and non-linear functions even of
the multipliers. There could then be irreducible higher-derivative terms of
multipliers that cannot be rewritten to be proportional to the underived
multiplier. However, such corrections could only appear if the multipliers
themselves were subject to quantization or other modifications, which never
happens in canonical approaches. The multipliers are not turned into operators
in canonical quantizations; they remain test functions even for constraint
operators. Moreover, they appear in classical constraints without their
spatial derivatives, so that they are not subject to discretization
modifications. It is therefore safe to assume that all terms in a given
effective constraint are proportional to one multiplier function without any
one of its derivatives.

\subsubsection{Derivative expansion of constraint brackets}
\label{s:DerivConstr}

For Poisson brackets of two smeared constraints, the previous considerations
take on a rather different form. As we will see explicitly below, if we assume
two constraints, $C_1[M]$ and $C_2[N]$, their Poisson bracket
$\{C_1[M],C_2[N]\}=\sum_{i,j} \int M^{(i)}N^{(j)} f_{i,j} {\rm d}x$ may depend
on higher spatial derivatives of $M$ and $N$, up to some order considered for
a derivative expansion of the constraints. The presence of two independent
functions $M$ and $N$ implies new features compared to the previous discussion
of a single constraint. First, it is, in general, no longer possible to remove
all spatial derivatives of $M$ and $N$ by integrating by parts in $\sum_{i,j}
\int M^{(i)}N^{(j)} f_{i,j} {\rm d}x$. Some higher spatial derivatives of
multipliers will therefore remain in Poisson brackets even if they can always
be removed in the constraints themselves. We may assume a form in which one of
the multipliers, say $M$, appears without its derivatives,
$\{C_1[M],C_2[N]\}=\sum_{j} \int MN^{(j)} g_{j} {\rm d}x$ with new functions
$g_j$, but trying to remove further the derivatives of $N$ will reinstate
those of $M$. As with an individual constraint, the latter form with underived
$M$ may be used to fix some of the freedom of integrating by parts, but it
will not remove all spatial derivatives of multipliers.

Secondly, and more importantly, the presence of two independent multiplier
functions implies that there are several independent terms in the Poisson
bracket of two constraints. If the bracket is required to have a certain form,
for instance that it be first class and therefore vanish on the constraint
surface, several independent conditions will result. To see this, we must
consider the freedom contained in a pair of functions, or the set
$\{(M,N): M,N\mbox{ functions
  on space}\}$. For a first-class algebra $\{C_1[M],C_2[N]\}$, we have the
condition that $\sum_{j} \int MN^{(j)} g_{j} {\rm d}x$ be a linear combination
of all original constraints. For a single multiplier in this expression, there
would be just one condition. With two multipliers $M$ and $N$, however, a new
condition arises for each derivative order $j$.

To show this, we work locally without loss of generality because it is
sufficient to vary functions in a neighborhood ${\cal U}$ of an arbitrary but
fixed point to derive equations of motion. Furthermore, we may assume the
multipliers to be smooth and Taylor-expandable in the chosen neighborhood. We
may then re-organize our set of local multiplier functions as
\begin{eqnarray}
&&\{(M,N):M,N\mbox{ smooth functions on }{\cal U}\}\\
&=& \left\langle\bigcup_j
  \{(M,N): M \mbox{ a smooth function on }{\cal U}\mbox{ and }N=x^j\}
\right\rangle \nonumber
\end{eqnarray}
using all monomials of degree $j$ for $N$, and denoting by
$\langle\cdot\rangle$ the linear span. We then derive iteratively that all
$g_j$ must independently be a combination of constraints: For $N=c_0$ constant
and varying by $c_0$, we have that $\int Mg_0{\rm d}x$ must be a combination
of constraints for all $M$, so that $g_0$ must locally be a combination of
constraints. For $N=c_1x$, varying by $c_1$ and using the first result on
$g_0$, we obtain that $\int Mg_1{\rm d}x$ must be a combination of
constraints, still for all $M$ since the $M$-variations of $(M,N)$ with
$N=c_1x$ are independent of those with $N=c_0$. Proceeding in this way, all
$g_j$ must independently be combinations of the constraints. A first-class
algebra of constraints with higher spatial derivatives therefore requires
additional conditions on the possible form of constraints, even if no
additional constraints on phase space are implied.

In the preceding argument on the independence of multiplier functions and
independent conditions $g_j$ it was important that the Poisson bracket
$\{C_1[M],C_2[N]\}$ was assumed to be arranged in the form $\sum_{j} \int
MN^{(j)} g_{j} {\rm d}x$, using integrations by parts. Sometimes, especially
for the bracket of two Hamiltonian constraints, the series may at first appear
in a different form. In the next subsection, we will see that for two
Hamiltonian constraints (or more generally, for the bracket of two copies of
the same constraint with different multipliers) it is often more natural to
write the bracket as $\{H[M],H[N]\}= \sum_{i,j} \int
(M^{(i)}N^{(j)}-M^{(j)}N^{(i)}) h_{i,j}{\rm d}x$ to make the antisymmetry in
$M$ and $N$ explicit. The sum may be assumed to be such that $i<j$, with $j$
ranging from zero to $n$ at $n$-th order.  However, it turns out that the
antisymmetric combinations $M^{(i)}N^{(j)}-M^{(j)}N^{(i)}$ cannot all be
varied independently of one another, and that the $h_{i,j}$ in a first-class
algebra need not be combinations of constraints independently for all $i$ and
$j$. To see this, it suffices to rewrite the first few orders of an
antisymmetric arrangement in terms of the standard form used before. (Such
formulas will be useful for later manipulations in explicit examples. We
include the general expressions at $n$th order in an appendix.)

At first order, integrating by parts and ignoring boundary terms, we have 
\begin{equation} \label{NM1}
 \int{\rm d}x (MN'-M'N)h_{0,1}= \int {\rm d}x (MNh_{0,1}'+2MN'h_{0,1})\,,
\end{equation}
both forms require the same condition for a first-class algebra, namely that
$g_1=2h_{0,1}\approx 0$ vanish on the constraint surface (which implies that
$g_0=h_{0,1}'$ vanishes on the same surface). At second maximal order, $j=2$,
we have two additional terms
\begin{equation}\label{NM2}
 \int{\rm d}x(MN''-M''N)h_{0,2}= -\int{\rm d}x(MNh_{0,2}''+2MN'h_{0,2}')\,,
\end{equation}
and
\begin{equation}\label{NM3}
\int{\rm d}x (M'N''-M''N')h_{1,2}= -\int{\rm
  d}x(MN'h_{1,2}''+3MN''h_{1,2}'+2MN'''h_{1,2})\,,
\end{equation}
so that adding (\ref{NM1}), (\ref{NM2}) and (\ref{NM3}) results in
\begin{align}
\int{\rm d}x\sum_{j=1}^2\sum_{i=0}^{j-1}(M^{(i)}N^{(j)}-M^{(j)}N^{(i)})h_{i,j}=\int & \md
x\,(-2MN'''h_{1,2}-3MN''h'_{1,2}\\ 
+&MN'(2h_{0,1}-2h'_{0,2}-h_{1,2}'') +MN(h'_{0,1}-h_{0,2}'')) \notag
\end{align}
giving four conditions
\begin{eqnarray*}
g_3=-2h_{1,2}&\approx&0\,,\\
g_2=-3h'_{1,2}&\approx&0\,,\\
g_1=2h_{0,1}-2h'_{0,2}-h_{1,2}''&\approx&0\,,\\
g_0=h'_{0,1}-h_{0,2}''&\approx&0\,,\\
\end{eqnarray*}
of which only two are independent:
\begin{equation}  \label{secondOrdCond2}
h_{1,2}\approx0\,,
\end{equation}
and
\begin{equation}  \label{secondOrdCond1}
h_{0,1}-h_{0,2}'\approx0\,.
\end{equation}
The functions $h_{0,1}$ and $h_{0,2}$ in (\ref{NM1}) and (\ref{NM2}) need not
vanish independently.

Similarly at third maximal order there are six conditions 
\begin{eqnarray*}
g_5=2h_{2,3}&\approx&0\,,\\
g_4=5h_{2,3}'&\approx&0\,,\\
g_3=-2h_{1,2}+2h_{0,3}+2h'_{1,3}+4h_{2,3}''&\approx&0\,,\\
g_2=-3h'_{1,2}+3h_{0,3}'+3h_{1,3}''+h_{2,3}'''&\approx&0\,,\\
g_1=2h_{0,1}-2h'_{0,2}-h_{1,2}''+3h_{0,3}''+h_{1,3}'''&\approx&0\,,\\
g_0=h'_{0,1}-h_{0,2}''+h_{0,3}'''&\approx&0\,,\\
\end{eqnarray*}
but only three of them are independent, implying:
\begin{eqnarray}
h_{2,3}&\approx&0\,, \label{hthree1} \\
h_{1,2}-h_{0,3}-h_{1,3}'&\approx&0\,, \label{hthree2} \\
h_{0,1}-h'_{0,2}+h_{0,3}''&\approx&0\,.  \label{hthree3}
\end{eqnarray}
As these examples indicate, even total orders $i+j$ do not lead to conditions
independent of those from odd orders because the highest even derivatives
$MN^{(i+j)}$ cancel out after integrating by parts the antisymmetric
combinations $M^{(i)}N^{(j)}-M^{(j)}N^{(i)}$.

\subsubsection{Functional dervatives}
 
We now turn to explicit formulas to compute Poisson brackets of constraints
with derivative corrections. These are again given for fields in one spatial
dimension (the case of interest for spherical symmetry), but most of them
easily generalize to higher spatial dimensions.

For a functional
\[
F[N,q]:=\int \md x N F(q(x),q'(x),q''(x),\dots,q^{(n)}(x))
\]
depending on some smearing function $N$, and a field $q$ and its spatial
derivatives up to order $n$, we compute its functional derivative using
\begin{equation}
\vd{F[N,q]}{q(x)}:=\vd{(NF)}{q}\bigg|_{q=q(x)}
\end{equation}
where $\delta (NF)/\delta q$ is the `variational derivative' of $NF$:
\begin{equation} \label{variationalDerivative}
\vd{(NF)}{q}:=\sum_{k=0}^{n}(-1)^k\left(N\Par{F}{q^{(k)}}\right)^{(k)}\,.
\end{equation}
Here and in what follows we use the same letter to denote the smeared  and
unsmeared (density) functional. 

We may expand terms of this form using the binomial identity for the $k$-th derivative of a product:
\begin{equation} \label{binomialIdentity}
(AB)^{(k)}=\sum_{l=0}^{k}\binom{k}{l}A^{(k-l)}B^{(l)}\,,
\end{equation}
and obtain
\begin{align}
\vd{(NF)}{q}&=\sum_{k=0}^{n}(-1)^k\left(N\Par{F}{q^{(k)}}\right)^{(k)} \notag\\
&=\sum_{k=0}^{n}\sum_{l=0}^{k}(-1)^k\binom{k}{l}N^{(k-l)}\left(\Par{F}{q^{(k)}}\right)^{(l)} \notag\\
&=\sum_{j=0}^{n}N^{(j)}\sum_{k=0}^{n-j}(-1)^{j+k}\binom{j+k}{k}\left(\Par{F}{q^{(j+k)}}\right)^{(k)}\,. \label{variationalDerivExpanded}
\end{align}
In the last line, we have used the identity:
\begin{equation} \label{PascalSum}
\sum_{k=0}^n\sum_{l=0}^k\binom{k}{l}A_{k,l}=\sum_{j=0}^n\sum_{k=0}^{n-j}\binom{j+k}{k}A_{j+k,k}\,,
\end{equation}
for arbitrary functions $A_{k,l}$.  
The right -hand side of (\ref{PascalSum})
follows from summing `diagonally' as opposed to row by row in the diagram:

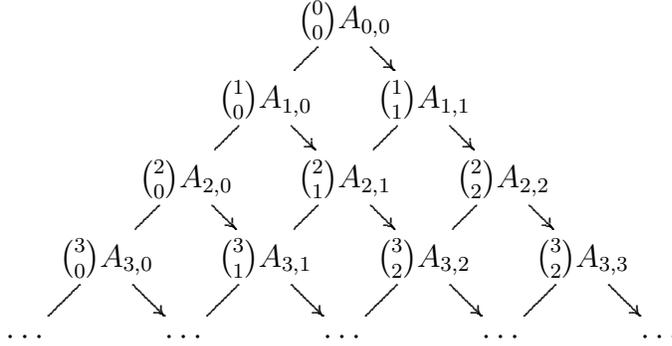
\begin{center}
\begindc{\commdiag} 
\obj(4,4)[n00]{${0 \choose 0} A_{0,0}$}
\obj(3,3)[n10]{${1 \choose 0} A_{1,0}$}
\obj(5,3)[n11]{${1 \choose 1} A_{1,1}$}
\obj(2,2)[n20]{${2 \choose 0} A_{2,0}$}
\obj(4,2)[n21]{${2 \choose 1} A_{2,1}$}
\obj(6,2)[n22]{${2 \choose 2} A_{2,2}$}
\obj(1,1)[n30]{${3 \choose 0} A_{3,0}$}
\obj(3,1)[n31]{${3 \choose 1} A_{3,1}$}
\obj(5,1)[n32]{${3 \choose 2} A_{3,2}$}
\obj(7,1)[n33]{${3 \choose 2} A_{3,3}$}
\obj(0,0)[n40]{$\cdots$}
\obj(2,0)[n41]{$\cdots$}
\obj(4,0)[n42]{$\cdots$}
\obj(6,0)[n43]{$\cdots$}
\obj(8,0)[n44]{$\cdots$}
\mor{n00}{n10}{}[1, \solidline]
\mor{n00}{n11}{}
\mor{n10}{n20}{}[1, \solidline]
\mor{n10}{n21}{}
\mor{n11}{n21}{}[1, \solidline]
\mor{n11}{n22}{}
\mor{n20}{n30}{}[1, \solidline]
\mor{n20}{n31}{}
\mor{n21}{n31}{}[1, \solidline]
\mor{n21}{n32}{}
\mor{n22}{n32}{}[1, \solidline]
\mor{n22}{n33}{}
\mor{n30}{n40}{}[1, \solidline]
\mor{n30}{n41}{}
\mor{n31}{n41}{}[1, \solidline]
\mor{n31}{n42}{}
\mor{n32}{n42}{}[1, \solidline]
\mor{n32}{n43}{}
\mor{n33}{n43}{}[1, \solidline]
\mor{n33}{n44}{}
\enddc
\end{center}

Making use of (\ref{variationalDerivExpanded}), we may now write, for general
funcionals $F_A[M,q,\dots,q^{(m)}]$ and $F_B[N,p,\dots,p^{(n)}]$, the formula:
\begin{equation} \label{FAFB}
\vd{(MF_A)}{q}\vd{(NF_B)}{p}-(M\leftrightarrow N)=\sum_{i=0}^{m}\sum_{\substack{j=0 \\ j\neq i}}^{n}\big(M^{(i)}N^{(j)}-N^{(i)}M^{(j)}\big)\delta_{q}^{m,i}F_A\,\delta_{p}^{n,j}F_B
\end{equation}
with
\begin{equation}
\delta_q^{m,i}F:=\sum_{k=0}^{m-i}(-1)^{i+k}\binom{i+k}{k}
\left(\Par{F}{q^{(i+k)}}\right)^{(k)} \,.
\end{equation}

These basic formulas equally apply to the brackets $\{H[N],H[M]\}$ and
$\{H[N],D[N^x]\}$. However, since we assume the diffeomorphism constraint $D[N^x]$ to be unaffected by
quantum corrections and therefore to remain of first order in spatial
derivatives, we may write the bracket containing it in more explicit
form. 

Returning to general expressions (\ref{variationalDerivative}), (\ref{binomialIdentity}), and  (\ref{PascalSum}), with the assumption of at most first
derivatives in $N^x$, from the form of (\ref{FDbracket}) for the Poisson bracket of a general functional $F[N]$ with  $D[N^x]$,  for each scalar variable $q$ there will then be a
contribution of the form
\begin{equation} 
\int \md x\, N^x\,\vd{(NF)}{q}q'
\end{equation}
and  (if $q$ is a density of weight one) an additional one of the form
\begin{equation} \label{densityContrib}
\int \md x\, (N^x)'\,\vd{(NF)}{q}q\,.
\end{equation}

Using (\ref{variationalDerivative}), (\ref{binomialIdentity}), and  (\ref{PascalSum}), with added integration by parts, we may write
\begin{align}
\int \md x\, N^x\,\vd{(NF)}{q}q'&=\int \md x\,N^x\,\sum_{k=0}^{n}(-1)^{k}\left(N\Par{F}{q^{(k)}}\right)^{(k)}q' \notag\\
&=\int \md x\,\sum_{k=0}^{n}(-1)^{2k}\left(N\Par{F}{q^{(k)}}\right)(N^xq')^{(k)} \notag\\
&=\int \md x\,\sum_{k=0}^{n}\sum_{l=0}^{k}N(N^x)^{(k-l)}\binom{k}{l}\Par{F}{q^{(k)}}q^{(l+1)} \notag\\
&=\int \md x\bigg[NN^x\sum_{k=0}^{n}\Par{F}{q^{(k)}}q^{(k+1)}+   \sum_{k=1}^{n}\sum_{l=0}^{k-1}N(N^x)^{(k-l)}\binom{k}{l}\Par{F}{q^{(k)}}q^{(l+1)}\bigg] \notag\\ \label{FDscalarContribution}
&=\int \md x\bigg[NN^x\sum_{k=0}^{n}\Par{F}{q^{(k)}}q^{(k+1)}+ \sum_{i=1}^{n}N(N^x)^{(i)}\sum_{k=0}^{n-i}\binom{i+k}{k}\Par{F}{q^{(i+k)}}q^{(k+1)}\bigg]\,. \notag\\
\end{align}
Similarly we can write
\begin{equation}
\int \md x\, (N^x)'\,\vd{(NF)}{q}q=\int \md x\,\sum_{i=1}^{n+1}N(N^x)^{(i)}\left(\Par{F}{q^{(i-1)}}q+\sum_{k=0}^{n-i}\binom{i+k}{k+1}\Par{F}{q^{(i+k)}}q^{(k+1)}\right)\,,
\end{equation}
so that the total contribution for a density is
\begin{align}
\int \md x\, \bigg[& N^x\,\vd{(NF)}{q}q' + (N^x)'\,\vd{(NF)}{q}q\bigg]=\int \md x\bigg[NN^x\sum_{k=0}^{n}\Par{F}{q^{(k)}}q^{(k+1)} \notag\\
&+ \sum_{i=1}^{n}N(N^x)^{(i)}\bigg(\Par{F}{q^{(i-1)}}q+\sum_{k=0}^{n-i}\binom{i+k+1}{k+1}\Par{F}{q^{(i+k)}}q^{(k+1)}\bigg) \notag\\
&+N(N^x)^{(n+1)}\Par{F}{q^{(n)}}q \,\bigg]\,.  \label{FDdensityContribution}
\end{align}
These expressions can readily be used to explicitly calculate the bracket of any phase space functional with the diffeomorphism constraint in one dimension.

To compute the Poisson bracket of two Hamiltonian constraints we may use (\ref{FAFB}), for the special case  $F_A=F_B=H$:
\begin{align} 
\vd{(MH)}{q}\vd{(NH)}{p}-(M\leftrightarrow N)&=\sum_{i=0}^{n}\sum_{\substack{j=0 \\ j\neq i}}^{n}\big(M^{(i)}N^{(j)}-N^{(i)}M^{(j)}\big)\delta_{q}^{n,i}H\,\delta_{p}^{n,j}H \notag\\
&=\sum_{j=1}^{n}\sum_{i=0}^{j-1} \big(M^{(i)}N^{(j)}-N^{(i)}M^{(j)}\big)\big(\delta_{q}^{n,i}H\,\delta_{p}^{n,j}H-\delta_{q}^{n,j}H\,\delta_{p}^{n,i}H\big),   \label{FAFA}
\end{align}
where now $n$ is the maximum derivative order considered of the two variables
$q$ and $p$. (This is true because for $m>n$, with $n$ the maximum order of
derivatives of $q$ appearing in $H$, we have
$\delta_{q}^{m,i}H=\delta_{q}^{n,i}H$ and $\delta_{q}^{s,m}H=0$.) Application
of this formula for each canonical pair gives rise to an expression of the
form
\begin{equation} \label{antisymHHbracket}
\{H[M],H[N]\}=\sum_{j=1}^n\sum_{i=0}^{j-1}\int \md
x\,(M^{(i)}N^{(j)}-N^{(i)}M^{(j)})h_{i,j} 
\end{equation}
alluded to previously. 

The appendix shows how one can use the lower order calculations from the
previous subsection to rewrite the general expression (\ref{antisymHHbracket})
in the form $\sum_{j=0}^{2n-1}\int \md x\, NM^{(j)}g_j$.

\subsection{Spherical symmetry}

We now specialize the previously derived formulas to the spherically symmetric case.
The Poisson bracket of functions $f$ and $g$ on the phase space of spherically
symmetric gravity is
\begin{align}
\{f,g\}=2G\int \md x\bigg(&\gamma\vd{f}{A_x}\vd{g}{E^x}+\frac{1}{2}\vd{f}{\Kf}\vd{g}{\Ef}+\gamma\vd{f}{\eta}\vd{g}{P^\eta}  \notag\\
&-\gamma\vd{f}{E^x}\vd{g}{A_x}-\frac{1}{2}\vd{f}{\Ef}\vd{g}{\Kf}-\gamma\vd{f}{P^\eta}\vd{g}{\eta}\bigg)\,, \label{AppendixPBracket}
\end{align}
which we apply to $f$ and $g$ being the Hamiltonian or diffeomorphism
constraints.  

The general modified Hamiltonian constraint we consider is:
\begin{align}
H[N]=-\frac{1}{2G}\int \md x\,N \big(& \alpha\,|E^x|^{-\frac{1}{2}}E^\varphi f_1+2s\bar{\alpha}\,|E^x|^\frac{1}{2}f_2 + \alpha\,|E^x|^{-\frac{1}{2}}E^\varphi   \notag \\
&-\alpha_\Gamma\,|E^x|^{-\frac{1}{2}}E^\varphi\Gamma_\varphi^2+ 2s\bar{\alpha}_\Gamma\,|E^x|^\frac{1}{2}\Gamma_\varphi'\big)\,,   \label{ModifiedHamiltonian}
\end{align}
where $s={\rm sign} E^x$. We use $\Gamma_{\varphi}=-(E^x)'/(2E^{\varphi})$ as
an abbreviation, which classically would be a component of the spin
connection. Classically, $f_1=\Kf^2$ and $f_2=\Kf(A_x+\eta')/\gamma$ for
spherically symmetric gravity, and
$\alpha=\bar{\alpha}=\alpha_\Gamma=\bar{\alpha}_\Gamma=1$. Not all these
functions are independent and we could, for instance, absorb $\alpha$ in
$f_1$. However, we will keep them separate to indicate their different origins
in inverse-triad and holonomy corrections, respectively.  

The Gauss and diffeomorphism constraints remain unaltered because their
classical action on phase space can directly be lifted to quantum states. The
gauge transformations they generate are therefore unmodified, and we have
\begin{equation}
\mathcal{G}[\lambda]=\frac{1}{2G\gamma}\int \md x\,
\lambda((E^x)'+P^\eta)  \label{GaussConstraint}
\end{equation}
\begin{equation}
D[N^x]=\frac{1}{2G}\int \md x\,N^x\left(2E^\varphi K_\varphi'-\frac{1}{\gamma}A_x(E^x)'+\frac{1}{\gamma}\eta'P^\eta\right) \label{DiffeoConstraint}
\end{equation}
We keep the full set of constraints, but one can easily solve the Gauss
constraint by replacing $A_x/\gamma$ with $K_x$ and eliminating $\eta'$
terms. (The extrinsic-curvature component $K_x=\gamma^{-1}(A_x+\eta')$ is
invariant under the action generated by the Gauss constraint, and $P^{\eta}$
is expressed in terms of $(E^x)'$ on its constraint surface.)

We first give general formulas to compute the Poisson algebra of constraints
when all the correction functions ($f_1$, $f_2$, $\alpha$, $\bar{\alpha}$,
$\alpha_\Gamma$ and $\bar{\alpha}_\Gamma$) are allowed to be arbitrary
(smooth) functions of the configuration variables $A_x+\eta'$, $\Kf$, triads
$E^x$, $\Ef$, and their derivatives to some order $n$. For a more detailed
analysis, we then specialize to the cases of $n=0$ and $n=1$ for
holonomy corrections with or without inverse triad corrections.
For these explicit considerations of Poisson brackets, we will find it convenient to split the
Hamiltonian constraint (\ref{ModifiedHamiltonian}) into its terms
\begin{align}
H_0[N]&:=-\frac{1}{2G}\int \md x\, N(\alpha\,|E^x|^{-\frac{1}{2}}\Ef( f_1 +1)) \notag\\
H_A[N]&:=-\frac{1}{2G}\int \md x\, N(2s\bar{\alpha}\,|E^x|^\frac{1}{2}f_2) \notag \\
H_\Gamma[N]&:=H_\Gamma^1[N]+H_\Gamma^2[N]+H_\Gamma^3[N] \notag
\end{align}
with
\begin{align}
H_\Gamma^1[N]&:=\frac{1}{2G}\int \md x\, N\,\alpha_\Gamma
\frac{|E^x|^{-\frac{1}{2}} ((E^x)')^2}{4\Ef}  \notag\\ 
H_\Gamma^2[N]&:=\frac{1}{2G}\int \md x\, N\,s\bar{\alpha}_\Gamma
\frac{|E^x|^{\frac{1}{2}} (E^x)''}{\Ef} \notag \\ 
H_\Gamma^3[N]&:=-\frac{1}{2G}\int \md x\, N\,s\bar{\alpha}_\Gamma
\frac{|E^x|^{\frac{1}{2}}(E^x)' (\Ef)'}{\Ef\,^2} \notag 
\end{align}

\subsubsection{Diffeomorphism bracket}
\label{s:Diffeo}

The bracket $\{F,D[N^x]\}$ of some phase-space function $F$ with
the diffeomorphism constraint may be computed explicitly for an arbitrary
dependence of $F$ on the canonical variables and their derivatives up to
$n$-th order, as indicated in the preceding subsection.  Given the form of the
diffeomorphism constraint (\ref{DiffeoConstraint}) we have
\begin{align}
\{F[N],D[N^x]\}=\int \md x N^x \bigg[&\vd{F[N]}{A_x}A_x'+\vd{F[N]}{\eta}\eta'+\vd{F[N]}{\Kf}\Kf' \notag\\
&+\vd{F[N]}{E^x}E^x\,'+\vd{F[N]}{\Ef}\Ef\,'+\vd{F[N]}{P^\eta}P^\eta\,'\bigg] \notag\\
+\int \md x
(N^x)'\,&\bigg[\vd{F[N]}{A_x}A_x+\vd{F[N]}{\Ef}\Ef+\vd{F[N]}{P^\eta}P^\eta\bigg] \,.  \label{FDbracket}
\end{align}

Specializing $F$ to the Hamiltonian constraint $H$, and using (\ref{FDscalarContribution}) and (\ref{FDdensityContribution}), the functional derivatives take the form

\begin{align}
\{H[N],D[N^x]&\}=\int \md x\,NN^x\,H'  \notag\\
+\sum_{i=1}^n\int \md x&\, N(N^x)^{(i)}\Bigg[\Par{H}{A_x^{(i-1)}}A_x+\Par{H}{\eta^{(i)}}\eta'+\Par{H}{(\Ef)^{(i-1)}}\Ef  \notag\\
+\sum_{k=0}^{n-i}\Bigg(&\binom{i+k+1}{k+1}\bigg(\Par{H}{A_x^{(i+k)}}A_x^{(k+1)}+\Par{H}{\eta^{(i+k+1)}}\eta^{(k+2)}\bigg)+\binom{i+k}{k}\Par{H}{\Kf^{(i+k)}}\Kf^{(k+1)} \notag\\
&+\binom{i+k}{k}\Par{H}{(E^x)^{(i+k)}}(E^x)^{(k+1)}\,+\,\binom{i+k+1}{k+1}\Par{H}{(\Ef)^{(i+k)}}(\Ef)^{(k+1)}\bigg)\Bigg] \notag\\
+\int \md x&\,
N(N^x)^{(n+1)}\Bigg[\Par{H}{A_x^{(n)}}A_x+\Par{H}{\eta^{(n+1)}}\eta'+\Par{H}{(\Ef)^{(n)}}\Ef\Bigg] \,,
\end{align}
where $n$ is the maximum order considered.

The explicit dependence of (\ref{ModifiedHamiltonian}) on $E^x$, $\Ef$, $(E^x)'$,
$(\Ef)'$ and $(E^x)''$ gives the contribution term
\[
\int \md x\,N(N^x)'\left(H_0+H_\Gamma\right)=\int \md x\, N(N^x)'\left(H-H_A\right)\,,
\]
so we may also write
\begin{align}
\{H[N],D[N^x]&\}=\int \md x\,NN^x\,H'+\int \md x\,N(N^x)'(H-H_A)  \notag\\
+\sum_{i=1}^n\int \md x&\, N(N^x)^{(i)}\Bigg[\Par{H}{A_x^{(i-1)}}A_x+\Par{H}{\eta^{(i)}}\eta'+\CPar{H}{(\Ef)^{(i-1)}}\Ef  \notag\\
+\sum_{k=0}^{n-i}\Bigg(&\binom{i+k+1}{k+1}\bigg(\Par{H}{A_x^{(i+k)}}A_x^{(k+1)}+\Par{H}{\eta^{(i+k+1)}}\eta^{(k+2)}\bigg)+\binom{i+k}{k}\Par{H}{\Kf^{(i+k)}}\Kf^{(k+1)} \notag\\
&+\binom{i+k}{k}\CPar{H}{(E^x)^{(i+k)}}(E^x)^{(k+1)}\,+\,\binom{i+k+1}{k+1}\CPar{H}{(\Ef)^{(i+k)}}(\Ef)^{(k+1)}\bigg)\Bigg] \notag\\
+\int \md x&\,
N(N^x)^{(n+1)}\Bigg[\Par{H}{A_x^{(n)}}A_x+\Par{H}{\eta^{(n+1)}}\eta'+\CPar{H}{(\Ef)^{(n)}}\Ef\Bigg] \,.
\label{HDBracket}
\end{align}
With the $i$-sum being zero for $n=0$.

Here and in what follows we use the short hand notation
\[
 \CPar{H}{q}
\]
for the partial derivative of $H$ with respect to $q$, acting only on the
correction functions. Notice that ${\rm D}H/{\rm D} A_x^{(i)}=\partial
H/\partial A_x^{(i)}$ for all $i\geq 0$, ${\rm D}H/{\rm D}(\Ef)^{(i)}=\partial
H/\partial (\Ef)^{(i)}$ for $i>1$, and ${\rm D}H/{\rm D}(E^x)^{(i)}=\partial
H/\partial (E^x)^{(i)}$ for $i>2$. We could therefore have written all terms
in the previous expression as well as ${\rm D}$-derivatives.
 
If we allow for one higher order of derivatives of the triad, and we group derivatives $A^{(k)}$ with derivatives $E^{(k+1)}$, we rearrange as
\begin{align}
\{H[N],D[N^x]\}=\int \md x\,N&N^x\,H'+\int \md x\,N(N^x)'(H-H_A)  \notag\\
+\sum_{i=1}^n\int \md x\, N(N^x)^{(i)}\Bigg[&\Par{H}{A_x^{(i-1)}}A_x+\Par{H}{\eta^{(i)}}\eta'+\CPar{H}{(\Ef)^{(i-1)}}\Ef \notag\\
&+  \CPar{H}{(E^x)^{(i)}}(E^x)'+(i+1)\CPar{H}{(\Ef)^{(i)}}(\Ef)'\notag\\
+\sum_{k=0}^{n-i}\Bigg(\binom{i+k+1}{k+1}&\bigg(\Par{H}{A_x^{(i+k)}}A_x^{(k+1)}+\Par{H}{\eta^{(i+k+1)}}\eta^{(k+2)}\bigg)+\binom{i+k}{k}\Par{H}{\Kf^{(i+k)}}\Kf^{(k+1)} \Bigg) \notag\\
+\sum_{k=1}^{n+1-i}\Bigg(\binom{i+k}{k}&\CPar{H}{(E^x)^{(i+k)}}(E^x)^{(k+1)}\,+\,\binom{i+k+1}{k+1}\CPar{H}{(\Ef)^{(i+k)}}(\Ef)^{(k+1)}\Bigg)\Bigg] \notag\\
+\int \md x\,N(N^x)^{(n+1)}\Bigg[&\Par{H}{A_x^{(n)}}A_x+\Par{H}{\eta^{(n+1)}}\eta'+\CPar{H}{(\Ef)^{(n)}}\Ef \notag\\
&+\CPar{H}{(E^x)^{(n+1)}}(E^x)'+(n+2)\CPar{H}{(\Ef)^{(n+1)}}(\Ef)'\Bigg] \notag\\
+\int \md x\,N(N^x)^{(n+2)}&\CPar{H}{(\Ef)^{(n+1)}}\Ef\,. \label{HDKnEnn}
\end{align}
The last integral immediately shows that the Hamiltonian does not depend on
$(\Ef)^{(n+1)}$.  There can only be non-trivial dependence on $(E^x)^{(n+1)}$.

\subsubsection{Hamiltonian bracket}

For $n_{A_x}$, $n_{E^x}$, $n_{\Kf}$, $n_{\Ef}$, the maximum order of derivatives of the corresponding variables, 
we compute the $\{H[M],H[N]\}$ bracket using  formula (\ref{FAFB})
\begin{align}
 \{H[M],H[N]\}=&2G\sum_{i=0}^{n_{A_x}}\sum_{\substack{j=0 \\ j\neq i}}^{n_{E^x}}\int \md x\,\big(M^{(i)}N^{(j)}-N^{(i)}M^{(j)}\big)\left[\gamma\,\delta_{A_x}^{n_{A_x},i}H\,\delta_{E^x}^{n_{E^x},j}H\right]+   \notag \\
 +&2G\sum_{i=0}^{n_{\Kf}}\sum_{\substack{j=0 \\ j\neq i}}^{n_{\Ef}}\int \md x\,\big(M^{(i)}N^{(j)}-N^{(i)}M^{(j)}\big)\left[\frac{1}{2}\delta_{\Kf}^{n_{\Kf},i}H\,\delta_{\Ef}^{n_{\Ef},j}H \right]\,,   \notag
\end{align}
or more concisely using (\ref{FAFA}) with $n=\max(n_{A_x}, n_{E^x}, n_{\Kf}, n_{\Ef})$:
\begin{align} 
 \{H[M],H[N]\}=2G\sum_{j=1}^{n}\sum_{i=0}^{j-1}\int \md x\,\big(M^{(i)}N^{(j)}-N^{(i)}M^{(j)}\big)\bigg[&\gamma(\delta_{A_x}^{n,i}H\,\delta_{E^x}^{n,j}H-\delta_{A_x}^{n,j}H\,\delta_{E^x}^{n,i}H) \notag\\
 &+\frac{1}{2}(\delta_{\Kf}^{n,i}H\,\delta_{\Ef}^{n,j}H-\delta_{\Kf}^{n,j}H\,\delta_{\Ef}^{n,i}H) \bigg] \,.
 \label{HHBracket}
\end{align}
where again
\begin{equation} \label{deltaH}
 \delta_{q}^{n,i}H:=\sum_{k=0}^{n-i}(-1)^{i+k}\binom{i+k}{k}\left(\Par{H}{q^{(i+k)}}\right)^{(k)}\,.
\end{equation}
For instance, at third maximal derivative order we have
\begin{equation}
 \delta_q^{3,0} H= \frac{\partial H}{\partial q}-\left(\frac{\partial
     H}{\partial q'}\right)'+ \left(\frac{\partial H}{\partial q''}\right)''-
 \left(\frac{\partial H}{\partial q'''}\right)'''
\end{equation}
and $\delta_q^{3,3}H=-\partial H/\partial q'''$. 

Taking into account the explicit dependence of $H$ on $E^x$, $\Ef$, $(E^x)'$,
$(\Ef)'$ and $(E^x)''$, we compute the coefficients in
(\ref{deltaH}). Defining
\begin{align}
 \Delta_{E^x}^0:=\frac{1}{2G}\bigg[&\frac{s}{2}\alpha|E^x|^{-\frac{3}{2}}
\Ef(f_1+1)-\bar{\alpha}|E^x|^{-\frac{1}{2}}f_2
-s\alpha_\Gamma\frac{|E^x|^{-\frac{3}{2}}((E^x)')^2}{8\Ef} \notag\\
 &+\bar{\alpha}_\Gamma\frac{|E^x|^{-\frac{1}{2}}(E^x)''}{2\Ef}
-\bar{\alpha}_\Gamma\frac{|E^x|^{-\frac{1}{2}}(E^x)'(\Ef)'}{2\Ef\,^2} \notag\\
 &-\bigg(\alpha_\Gamma\frac{|E^x|^{-\frac{1}{2}}(E^x)'}{2\Ef}\bigg)'+
\bigg(s\bar{\alpha}_\Gamma\frac{|E^x|^{\frac{1}{2}}(\Ef)'}{\Ef\,^2}\bigg)' 
+\bigg(s\bar{\alpha}_\Gamma\frac{|E^x|^{\frac{1}{2}}}{\Ef}\bigg)''\bigg] \\
 \Delta_{E^x}^1:=\frac{1}{2G}\bigg[&-\alpha_\Gamma
\frac{|E^x|^{-\frac{1}{2}}(E^x)'}{2\Ef}+
s\bar{\alpha}_\Gamma\frac{|E^x|^{\frac{1}{2}}(\Ef)'}{\Ef\,^2}+
2\bigg(s\bar{\alpha}_\Gamma\frac{|E^x|^{\frac{1}{2}}}{\Ef}\bigg)'\bigg] \\
 \Delta_{E^x}^2:=\frac{1}{2G}
\bigg(&s\bar{\alpha}_\Gamma\frac{|E^x|^{\frac{1}{2}}}{\Ef}\bigg) \\ 
 \Delta_{\Ef}^0:=\frac{1}{2G}\bigg[&\alpha|E^x|^{-\frac{1}{2}}(f_1+1)-
\alpha_\Gamma\frac{|E^x|^{-\frac{1}{2}}((E^x)')^2}{4\Ef\,^2}-
s\bar{\alpha}_\Gamma\frac{|E^x|^{\frac{1}{2}}(E^x)''}{\Ef\,^2}
 \notag\\ 
 &+2s\bar{\alpha}_\Gamma\frac{|E^x|^{\frac{1}{2}}(E^x)'(\Ef)'}{\Ef\,^3} 
+\bigg(s\bar{\alpha}_\Gamma\frac{|E^x|^{\frac{1}{2}}(E^x)'}{\Ef\,^2}\bigg)'\bigg] \\
 \Delta_{\Ef}^1:=\frac{1}{2G}\bigg(&s\bar{\alpha}_\Gamma
\frac{|E^x|^{\frac{1}{2}}(E^x)'}{\Ef\,^2}\bigg),
\end{align}
we have
\begin{equation}
 \delta_{E^x}^{n,i}H:=\Delta_{E^x}^i+\sum_{k=0}^{n-i}(-1)^{i+k}\binom{i+k}{k}\left(\CPar{H}{(E^x)^{(i+k)}}\right)^{(k)} 
\end{equation}
for  $i=0,1,2$, and
\begin{equation}
 \delta_{\Ef}^{n,i}H:=\Delta_{\Ef}^i+\sum_{k=0}^{n-i}(-1)^{i+k}\binom{i+k}{k}\left(\CPar{H}{(\Ef)^{(i+k)}}\right)^{(k)} 
\end{equation}
for $i=0,1$.

\section{Examples}

A general treatment of closed constraint algebras in a derivative expansion
appears to be complicated, but one can deal with the lowest orders. At first
order (without additional derivatives beyond the classical form), we reproduce
but also strengthen the results of \cite{JR}. At second order, we will obtain
the first indications about possible higher-derivative corrections.

\subsection{No additional derivatives}

The case of an $H[N]$ with a modified dependence on phase-space variables but
no additional spatial derivatives has already been studied in \cite{JR}. However,
starting with more general assumptions on the possible dependence on $A_x$, we
will be able to strengthen previous results. It turns out that a consistent
deformation is possible with higher powers of $K_{\varphi}$. According to a
derivative expansion, one could expect terms with an additional spatial
derivative of $E$ to appear for each new factor of $K_{\varphi}$ in a series
expansion of $f_1$, which will be discussed in the next subsection.

The bracket (\ref{HDBracket}) with $n=0$ reads
\begin{align}
\{H[N],D[N^x]\}&=-H[N^xN'] \notag\\
&+\int \md x\, N(N^x)'\left[\,
  -H_A+\Par{H}{A_x}A_x+\Par{H}{\eta'}\eta'+\CPar{H}{\Ef}\Ef\right]  
\end{align}
and,  for a first-class algebra, gives the condition
\begin{equation} \label{HDBracketCond00}
-H_A+\Par{H}{(A_x+\eta')}(A_x+\eta')+\CPar{H}{\Ef}\Ef
=\mathcal{F}_1H+\mathcal{F}_2D
\end{equation}
with some functions $\mathcal{F}_1$ and $\mathcal{F}_2$.

Since $H$, by assumption, does not contain derivatives of $\Kf$ we must have
$\mathcal{F}_2=0$. Explicitly, (\ref{HDBracketCond00}) then reads
\begin{align}
&\frac{1}{\alpha(f_1+1)}\bigg(\Par{(\alpha(f_1+1))}{(A_x+\eta')}(A_x+\eta')+\Par{(\alpha(f_1+1))}{\Ef}\Ef\bigg)H_0\notag\\
&+\frac{1}{\bar{\alpha}f_2}\bigg(-\bar{\alpha}f_2+\Par{(\bar{\alpha}f_2)}{(A_x+\eta')}(A_x+\eta')+\Par{(\bar{\alpha}f_2)}{\Ef}\Ef\bigg)H_A \notag\\
&+\frac{1}{\alpha_\Gamma}\bigg(\Par{\alpha_\Gamma}{(A_x+\eta')}(A_x+\eta')+\Par{\alpha_\Gamma}{\Ef}\Ef\bigg)H_\Gamma^1 \notag\\
&+\frac{1}{\bar{\alpha}_\Gamma}\bigg(\Par{\bar{\alpha}_\Gamma}{(A_x+\eta')}(A_x+\eta')+\Par{\bar{\alpha}_\Gamma}{\Ef}\Ef\bigg)H_\Gamma^{2,3}=\mathcal{F}_1H\,. \label{HDBracketCond0e}
\end{align}
Only the last two terms contain derivatives $(E^x)'$, $(\Ef)'$ and $(E^x)''$,
and therefore
\begin{align}
\mathcal{F}_1&=\frac{1}{\alpha_\Gamma}\bigg(\Par{\alpha_\Gamma}{(A_x+\eta')}(A_x+\eta')+\Par{\alpha_\Gamma}{\Ef}\Ef\bigg) \notag\\
&=\frac{1}{\bar{\alpha}_\Gamma}\bigg(\Par{\bar{\alpha}_\Gamma}{(A_x+\eta')}(A_x+\eta')+\Par{\bar{\alpha}_\Gamma}{\Ef}\Ef\bigg) \label{F1}
\end{align}
must be satisfied.

If there are no inverse-triad corrections, that is,
$\alpha=\bar{\alpha}=\alpha_\Gamma=\bar{\alpha}_\Gamma=1$, then (\ref{F1})
implies $\mathcal{F}_1=0$ and equation (\ref{HDBracketCond0e}) reads
\begin{equation} \label{HDCond0}
\Par{(H_0+H_A)}{(A_x+\eta')}(A_x+\eta')+\Par{(H_0+H_A)}{\Ef}\Ef=H_0+H_A
\end{equation}
with general solution\footnote{Define $h=\log(H_0+H_A)$, $y=\log(A_x+\eta')$
  and $z=\log E^{\varphi}$, and subsequently $y=Y+Z$ and $z=Y-Z$. The
  differential equation then reads $1=\partial h/\partial y+\partial
  h/\partial z=\partial h/\partial Y$, solved by $h(Y,Z)=Y+g(Z)$ with an
  arbitrary function $g(Z)$. In terms of the original variables, $H_0+H_A=
  \exp(g)\exp(Y)= \sqrt{(A_x+\eta')E^{\varphi}}G((A_x+\eta')/E^{\varphi})$. In
  the solution used in the text, we have, without restriction, rewritten
  $G(Z)=c_1 \exp(Z)+c_2\exp(-Z)+F(Z)$ because the first two terms appear in
  the classical constraint.}
\[
H_0+H_A=c_1\Ef+c_2(A_x+\eta')+F[(A_x+\eta')/\Ef]
\]
for functions $c_1$ and $c_2$ independent of $A_x$ and $\Ef$, and an arbitrary
function $F$. (If $f_1$ is assumed to be independent of $A_x+\eta'$, we have
the same equation and general solution for $f_2$.) We may discard the
homogeneous solution $F[(A_x+\eta')/\Ef]$ on the basis that $H_0+H_A$ has to
be of density weight one. Indeed, using the $\{H,H\}$ bracket we will see
explicitly that the dependence of $H$ on $A_x+\eta'$ has to be linear in this
case.

For general inverse-triad corrections, and if we assume all correction
functions except $f_2$ to be independent of $A_x+\eta'$, we get, by equating
the $H_A$ terms on the left and right hand side of (\ref{HDBracketCond0e}),
the slightly more complicated equation
\[
\Par{f_2}{(A_x+\eta')}(A_x+\eta')+\Par{f_2}{\Ef}\Ef=\bigg(\frac{1}{\alpha_\Gamma}\Par{\alpha_\Gamma}{\Ef}\Ef-\frac{1}{\bar{\alpha}}\Par{\bar{\alpha}}{\Ef}\Ef+1\bigg)f_2 \,.
 \]
It can be solved as before, with additional factors of derivatives of $\alpha$.

The $\{H[M],H[N]\}$ bracket (\ref{HHBracket}) is
\begin{align}
 \{H[M],H[N]\}=2G&\int \md x(MN'-NM')\\
& \times\bigg[-\gamma\Delta_{E^x}^2\left(\Par{H}{A_x}\right)'+\left(\Delta_{E^x}^1-(\Delta_{E^x}^2)'\right)\gamma\Par{H}{A_x}+\frac{1}{2}\left(\Delta_{\Ef}^1\right)\Par{H}{\Kf}\bigg]\,. \notag
\end{align}
 
Explicitly, for inverse-triad corrections independent of $A_x$,
\begin{align}
 \{H[M]&,H[N]\}=\label{HHBracket0}\\
=\frac{1}{2G}&\int \md x\,(MN'-NM')\Bigg[\;\bar{\alpha}\bar{\alpha}_\Gamma\frac{|E^x|}{\Ef\,^2}\left(2\gamma\left(\frac{\partial f_2}{\partial A_x}\right)'\Ef-\frac{\partial f_2}{\partial \Kf}(E^x)'\right)   \notag\\ 
+2\bigg(\bar{\alpha}_\Gamma\bar{\alpha}'&-\bar{\alpha}\bar{\alpha}_\Gamma'\bigg)\frac{|E^x|}{\Ef}\gamma\frac{\partial f_2}{\partial A_x}+s\bigg(\bar{\alpha}\alpha_\Gamma\gamma\frac{\partial f_2}{\partial A_x}-\alpha\bar{\alpha}_\Gamma\frac{1}{2}\frac{\partial f_1}{\partial \Kf}\bigg)\frac{(E^x)'}{\Ef} \notag\\
+s\alpha\bar{\alpha}_\Gamma\gamma\bigg(\frac{\partial f_1}{\partial A_x}\bigg)'&+s\bigg((\alpha\bar{\alpha}_\Gamma)'-2\alpha\bar{\alpha}_\Gamma'+\alpha\alpha_\Gamma\frac{(E^x)'}{2E^x}-\alpha\bar{\alpha}_\Gamma\bigg(\frac{(E^x)'}{E^x}-\frac{(\Ef)'}{\Ef}\bigg)\bigg)\gamma\frac{\partial f_1}{\partial A_x}\Bigg] \notag
\end{align}
 
Since inverse-triad corrections have already been studied in detail elsewhere,
we now consider holonomy corrections only, that is use
$\alpha=\bar{\alpha}=\alpha_\Gamma=\bar{\alpha}_\Gamma=1$. Condition
(\ref{HDCond0}) reads explicitly
 \begin{equation} \label{HDCond2}
 f_2-\frac{\partial f_2}{\partial \Ef}\Ef-\frac{\partial f_2}{\partial A_x}A_x
-\frac{\partial f_2}{\partial \eta'}\eta'
=\frac{\Ef}{2E^x}\left(\frac{\partial f_1}{\partial \Ef}\Ef
+\frac{\partial f_1}{\partial A_x}A_x
+\frac{\partial f_1}{\partial \eta'}\eta'\right)\,,
 \end{equation}
and the bracket (\ref{HHBracket0}) gives
\begin{align}
\{H[M]&,H[N]\}=\notag\\
=\frac{1}{2G}&\int \md x\,(MN'-NM')\Bigg[\;\frac{|E^x|}{(\Ef)^2}
\left(2\gamma\left(\frac{\partial
    f_2}{\partial A_x}\right)'\Ef-
\frac{\partial f_2}{\partial \Kf}(E^x)'\right)   \notag\\
+s\bigg(\gamma&\frac{\partial f_2}{\partial A_x}-
\frac{1}{2}\frac{\partial f_1}{\partial \Kf}\bigg)\frac{(E^x)'}{\Ef}+
s\gamma\bigg(\frac{\partial f_1}{\partial A_x}\bigg)'-
s\bigg(\frac{(E^x)'}{2E^x}-
\frac{(\Ef)'}{\Ef}\bigg)\gamma\frac{\partial f_1}{\partial A_x}\Bigg]\,.
\end{align}

Expanding 
\[
\left(\frac{\partial f_1}{\partial A_x}\right)'
=\frac{\partial^2 f_1}{\partial A_x^2}A_x'
+\frac{\partial^2 f_1}{\partial \Kf \partial A_x}\Kf'
+\frac{\partial^2 f_1}{\partial \eta' \partial A_x}\eta''
+\frac{\partial^2 f_1}{\partial E^x\partial A_x}(E^x)'
+\frac{\partial^2 f_1}{\partial \Ef \partial A_x}(\Ef)'
\]
and similarly for $\left(\partial f_2/\partial A_x\right)'$, we get
\begin{align}
\{&H[M],H[N]\}= \notag\\
&\frac{1}{2G}\int \md x\,(MN'-NM')\Bigg[\;\gamma\left(\frac{2|E^x|}{\Ef}\frac{\partial^2 f_2}{\partial \Kf \partial A_x}+s\frac{\partial^2 f_1}{\partial \Kf \partial A_x}\right)\Kf' \notag\\
&+\bigg(2\gamma\frac{|E^x|}{\Ef}\frac{\partial^2 f_2}{\partial E^x \partial A_x}+\frac{s\gamma}{\Ef}\frac{\partial f_2}{\partial A_x}-\frac{|E^x|}{\Ef\,^2}\frac{\partial f_2}{\partial \Kf}+s\gamma\frac{\partial^2 f_1}{\partial E^x \partial A_x}-\frac{\gamma}{2|E^x|}\frac{\partial f_1}{\partial A_x}-\frac{s}{2\Ef}\frac{\partial f_1}{\partial \Kf}\bigg)(E^x)' \notag\\
&+\bigg(2\gamma\frac{|E^x|}{\Ef}\frac{\partial^2 f_2}{\partial A_x^2}+s\gamma\frac{\partial^2 f_1}{\partial A_x^2}\bigg)A_x'+\bigg(2\gamma\frac{|E^x|}{\Ef}\frac{\partial^2 f_2}{\partial \eta'\partial A_x}+s\gamma\frac{\partial^2 f_1}{\partial \eta'\partial A_x}\bigg)\eta'' \notag\\
&+\bigg(2\gamma\frac{|E^x|}{\Ef}\frac{\partial^2 f_2}{\partial \Ef \partial A_x}+s\gamma\frac{\partial^2 f_1}{\partial \Ef\partial A_x}+\frac{s\gamma}{\Ef}\frac{\partial f_1}{\partial A_x}\bigg)(\Ef)' \Bigg] \,.  \label{HHBracketExpanded1}
\end{align}

If we impose again that the right hand side be a linear combination
$\mathcal{F}_1H+\mathcal{F}_2D$, by considering the $H^2_\Gamma$ term, we must
have $\mathcal{F}_1=0$ since correction functions do not contain second
derivatives of $E^x$. In order to have now a multiple of the diffeomorphism
constraint $2\Ef\Kf'-\frac{1}{\gamma}(A_x+\eta')(E^x)'$, the $A_x'$, $\eta''$
and $(\Ef)'$ terms must vanish:
\begin{align}
\frac{2E^x}{\Ef}\frac{\partial^2 f_2}{\partial A_x^2}+\frac{\partial^2 f_1}{\partial A_x^2}&=0 \label{HHCond1}\\
\frac{2E^x}{\Ef}\frac{\partial^2 f_2}{\partial \eta'\partial A_x}+\frac{\partial^2 f_1}{\partial \eta'\partial A_x}&=0 \label{HHCond2}\\
\frac{2E^x}{\Ef}\frac{\partial^2 f_2}{\partial \Ef \partial
  A_x}+\frac{\partial^2 f_1}{\partial \Ef\partial
  A_x}+\frac{1}{\Ef}\frac{\partial f_1}{\partial A_x}&=0 \,. \label{HHCond3}
\end{align}
If these conditions are not satisfied, the $A_x'$ and $(E^{\varphi})'$ terms
in (\ref{HHBracketExpanded1}) could be part of a first-class algebra only if
$\mathcal{F}_2$ depends on $A_x'$ and $(E^{\varphi})'$, respectively. Then we
would need $f_1$ or $f_2$ to depend on $A_x'$ and $(E^{\varphi})'$ for the first
terms in (\ref{HHBracketExpanded1}) to be anomaly-free with the same
dependence of $\mathcal{F}_2$ on derivatives, but such a dependence is assumed to be
absent in this subsection.

Equations (\ref{HHCond1}) and (\ref{HHCond2}) imply
\begin{equation} \label{HHCond12}
\frac{2E^x}{\Ef}\frac{\partial f_2}{\partial A_x}+\frac{\partial f_1}{\partial A_x}=C[\Kf,E^x,\Ef]
\end{equation}
or equivalently
\[
\frac{2E^x}{\Ef}f_2+f_1=C[\Kf,E^x,\Ef](A_x+\eta')+C_1[\Kf,E^x,\Ef]
\]
for arbitrary functions $C[\Kf,E^x,\Ef]$ and $C_1[\Kf,E^x,\Ef]$. Therefore,
under the present assumption, the Hamiltonian $H$ can depend only linearly
on $A_x+\eta'$. (From the equations, $C_1$ could also depend on $\eta'$, but
we know that $H$ must depend on the gauge invariant combination $A_x+\eta'$
only).

Substituting the derivative of equation (\ref{HHCond12}) with respect to $\Ef$
and (\ref{HHCond3}) back in (\ref{HHCond12}) gives the functional dependence
of $C$ on $\Ef$:
\[
\frac{\partial C}{\partial\Ef}=-\frac{C}{\Ef}
\]
so $C[\Kf,E^x,\Ef]=C_2[\Kf,E^x]/\Ef$, for some function $C_2[\Kf,E^x]$, and
\begin{equation} \label{linearAxDependence}
\frac{2E^x}{\Ef}f_2+f_1=C_2[\Kf,E^x]\frac{A_x+\eta'}{\Ef}+C_1[\Kf,E^x,\Ef]\,.
\end{equation}

Putting these results back in the bracket (\ref{HHBracketExpanded1})
\begin{align}
\{H[M],&H[N]\}= \notag\\
\frac{1}{2G}\int \md x\,&(MN'-NM')\Bigg[\;\frac{s\gamma}{2\Ef\,^2}\frac{\partial C_2}{\partial\Kf}\left(2\Ef\Kf'-\frac{1}{\gamma}(A_x+\eta')(E^x)'\right) \notag\\
&+s\gamma\left(\frac{\partial C_2}{\partial E^x}-\frac{C_2}{2E^x}-\frac{1}{2\gamma}\frac{\partial C_1}{\partial\Kf}\right)\frac{(E^x)'}{\Ef} \Bigg]
\end{align}
gives a condition for functions $C_1$ and $C_2$:
\[
\frac{\partial C_2}{\partial
  E^x}-\frac{C_2}{2E^x}-\frac{1}{2\gamma}\frac{\partial C_1}{\partial\Kf}=0 \,.
\]
Condition (\ref{HDCond2})  from the $\{H,D\}$ bracket translates into
\[
\frac{\partial C_1}{\partial\Ef}\Ef=0
\]
that is, $C_1$ is independent of $\Ef$.  Substituting
(\ref{linearAxDependence}) in the Hamiltonian then gives
\begin{align}
H[N]=-\frac{1}{2G}\int \md x\,N\big(& \,|E^x|^{-\frac{1}{2}}E^\varphi C_1+|E^x|^{-\frac{1}{2}}C_2(A_x+\eta')+|E^x|^{-\frac{1}{2}}E^\varphi   \notag \\
&-|E^x|^{-\frac{1}{2}}E^\varphi\Gamma_\varphi^2+ 2s|E^x|^\frac{1}{2}\Gamma_\varphi'\big) \notag
\end{align}
so changing $C\to C/\gamma$, $C_2\to C_2/\gamma$ and renaming variables gives
the general solution for $f_1$ and $f_2$ consistent with previous results:
\begin{align}
f_1&=C_1[\Kf,E^x]  \notag\\
f_2&=\frac{1}{2E^x}C_2[\Kf,E^x](A_x+\eta')/\gamma 
\end{align}
with $C_1$ and $C_2$ satisfying
\begin{equation} \label{C1C2}
 \frac{\partial C_1}{\partial \Kf}=2\frac{\partial C_2}{\partial
   E^x}-\frac{C_2}{E^x} \,.
\end{equation}

The form $f_1=C_1[\Kf,E^x]$ allows holonomy corrections to depend on the triad
component $E^x$ as well as on extrinsic curvature or the connection, which
could be used to model lattice refinement by a triad dependent length
parameter $\ell_0$. However, the relationship (\ref{C1C2}) rules out this kind
of parameterization: In a U(1)-theory as it automatically appears in the
reduced setting of spherical symmetry, the $K_{\varphi}$-dependence of
holonomies is almost-periodic. (On the quantum configuration space,
$K_{\varphi}$ takes values in the Bohr compactification of the real line.) If
we try to model lattice refinement by some form $\exp(if(E^x)K_{\varphi})$ of
holonomies, only a constant $f$ is compatible with (\ref{C1C2}) and an
almost-periodic $C_2$. Otherwise, the derivative
$\partial\exp(if(E^x)K_{\varphi})/\partial E^x=iK_{\varphi}({\rm d}f/{\rm
  d}E^x) \exp(if(E^x)K_{\varphi})$ is not almost periodic in
$K_{\varphi}$. Lattice refinement appears to be incompatible with a consistent
algebra if one insists on almost-periodic holonomy modifications. This result
shows an interesting relationship with problems of the Bohr compactification
as a model for non-Abelian connections, pointed out in \cite{NonAb}: Taking
into account the non-Abelian structure leads to a more-complicated
representation which automatically incorporates lattice refinement but is not
based on almost-periodic functions. (In isotropic models one can formally
write lattice-refined holonomies with $f$ of power-law form as unrefined ones
in variables redefined by a canonical transformation. This is not possible
here because we are dealing with two variables $K_{\varphi}$ and $E^x$ that
are not part of a canonical pair. The situation is closer to anisotropic
models, in which rescalings are not possible in general
\cite{SchwarzN}.)
 
The deformed algebra in this case is
\begin{align}
 \{H[N],D[N^x]\}=&-H[N^xN']  \notag\\
 \{H[M],H[N]\}=&\,D\left[(MN'-NM')\frac{s}{2(\Ef)^2}\frac{\partial C_2}{\partial \Kf}\right] \notag\\
 -G\bigg[(MN'&-NM')\eta'\frac{s}{2(\Ef)^2}\frac{\partial C_2}{\partial \Kf}\bigg]\,. \notag
\end{align}
The deformation function $\beta=(2E^x)^{-1}\partial C_2/\partial K_{\varphi}$,
which would equal one classically, is of particular interest. From
(\ref{C1C2}) we obtain 
\begin{equation} \label{betaC1}
 4\frac{\partial(E^x\beta)}{\partial E^x}-2\beta
 =2\beta-4E^x\frac{\partial\beta}{\partial E^x}= \frac{\partial^2C_1}{\partial
   K_{\varphi}^2}\,.
\end{equation}
If $\beta$ depends only weakly on $E^x$, which is expected for pure holonomy
corrections, it is negative near a maximum of $C_1$. As observed in
\cite{Action}, this behavior implies signature change to a quantum version of
4-dimensional Euclidean space whenever holonomy corrections are strong, in a
regime where they would bound the curvature dependence of the Hamiltonian
constraint.

These results agree with previous constructions, but they are more general
because we did not assume but rather derive that the Hamiltonian constraint
must depend linearly on $A_x+\eta'$ to the given order of derivatives.
 
\subsection{Higher spatial triad derivatives}

If we allow for higher spatial derivatives of the triad for extra factors of
$K_{\varphi}$ in an expanded $f_1$, as suggested by a derivative expansion,
several new terms appear in the equations of the previous subsection. For
instance, allowing for one additional order of spatial triad derivatives,
using (\ref{HDKnEnn}), the $\{H,D\}$ bracket and Eq.~(\ref{HDBracketCond00})
could include two additional terms $({\rm D}H/{\rm D}(E^x)') (E^x)'+ 2({\rm
  D}H/{\rm D}(E^{\varphi})') (E^{\varphi})'$. (The last one, however, must be
zero from the $N(N^x)''$-condition found in Sec.~\ref{s:Diffeo}.) Such terms
would be added to the explicit version (\ref{HDBracketCond0e}) as well. 

If one considers the implicit time derivative in $K_{\varphi}$ and $A_x$ on
the same footing as explicit higher spatial derivatives of the triad, a
consistent derivatives expansion must limit the polynomial order in
$K_{\varphi}$ and $A_x$ along with the explicit derivative order. If the next
derivative order is considered, with correction functions allowed to depend on
$(E^x)'$ and $(E^{\varphi})'$, the curvature-dependent functions $f_1$ and
$f_2$ could be third-order polynomials of $K_{\varphi}$ and $A_x$ or quadratic
with one spatial derivative. We are no longer allowed to assume a
non-polynomial function or a series for the curvature dependence, such as an
almost-periodic function. A third-order term in $K_{\varphi}$ or $A_x$ would
violate time-reversal symmetry, and so the next derivative order amounts to
correction functions depending on the first spatial derivative of the
phase-space variables. We will discuss such consistent versions in the next
subsection, and an explicit derivative expansions in more detail in
Sec.~\ref{s:ExtExp}.

\subsection{Dependence on first curvature derivatives}
 
We now consider the case where all correction function may depend on first
derivatives of the phase space variables. Bracket (\ref{HDBracket}) reads
accordingly
\begin{align}
\{H[N],D[N^x]\}=&-H[N^xN'] \notag\\
+\int \md x\, N&(N^x)'\bigg[\,
-H_A+\Par{H}{A_x}A_x+\Par{H}{\eta'}\eta'+\CPar{H}{\Ef}\Ef \notag\\ 
+2&\bigg(\Par{H}{A_x'}{A_x'}+\Par{H}{\eta''}\eta''\bigg)+\Par{H}{\Kf'}\Kf'+\CPar{H}{(E^x)'}(E^x)'+2\CPar{H}{(\Ef)'}(\Ef)'\bigg]  \notag\\
+\int \md x\, N&(N^x)''\bigg[\Par{H}{A_x'}A_x+\Par{H}{\eta''}\eta'+\CPar{H}{(\Ef)'}\Ef\bigg]\,. \label{HDBracket1}
\end{align}
Requiring the last integral proportional to $N(N^x)''$ to vanish weakly, gives
one condition, analogous to (\ref{HDBracketCond00})
\begin{equation} \label{HDBracketCond1}
\Par{H}{(A_x+\eta')'}(A_x+\eta')+\CPar{H}{(\Ef)'}\Ef
=\mathcal{F}_1H+\mathcal{F}_2D. 
\end{equation}

Again, since we assume unintegrated point holonomies for the
$\varphi$-components if we do not consider derivatives of $\Kf$ (dropping
$\partial H/\partial K_{\varphi}'$ in (\ref{HDBracket1})), we must have
$\mathcal{F}_2=0$. Explicitly,
\begin{align}
&\frac{1}{\alpha(f_1+1)}\bigg(\Par{(\alpha(f_1+1))}{(A_x+\eta')'}(A_x+\eta')+\Par{(\alpha(f_1+1))}{(\Ef)'}\Ef\bigg)H_0\notag\\
&+\frac{1}{\bar{\alpha}f_2}\bigg(\Par{(\bar{\alpha}f_2)}{(A_x+\eta')'}(A_x+\eta')+\Par{(\bar{\alpha}f_2)}{(\Ef)'}\Ef\bigg)H_A \notag\\
&+\frac{1}{\alpha_\Gamma}\bigg(\Par{\alpha_\Gamma}{(A_x+\eta')'}(A_x+\eta')+\Par{\alpha_\Gamma}{(\Ef)'}\Ef\bigg)H_\Gamma^1 \notag\\
&+\frac{1}{\bar{\alpha}_\Gamma}\bigg(\Par{\bar{\alpha}_\Gamma}{(A_x+\eta')'}(A_x+\eta')+\Par{\bar{\alpha}_\Gamma}{(\Ef)'}\Ef\bigg)H_\Gamma^{2,3}=\mathcal{F}_1H. \label{HDBracketCond1e}
\end{align}
Since only the last term contains the derivative $(E^x)''$, one must have
\begin{equation}
\mathcal{F}_1=\frac{1}{\bar{\alpha}_\Gamma}\bigg(\Par{\bar{\alpha}_\Gamma}{(A_x+\eta')}(A_x+\eta')+\Par{\bar{\alpha}_\Gamma}{\Ef}\Ef\bigg)
\end{equation}
Based on this equation we can already draw one conclusion: If there are no
inverse-triad corrections and no further corrections with derivatives of
$E^{\varphi}$ and $\Kf$, ${\cal F}_1=0$ and (\ref{HDBracketCond1}) implies
that there can be no dependence on $A_x'$ either.

With or without the former assumptions, the simplest possibility is for each
coefficient of $H_0$, $H_A$ and $H_\Gamma$ above to be equal to zero, so the
condition for the correction functions $\alpha(f_1+1)$, $\bar{\alpha}f_2$,
$\alpha_\Gamma$ and $\bar{\alpha}_\Gamma$ is
\[
\Par{F}{(A_x+\eta')'}(A_x+\eta')+\Par{F}{(\Ef)'}\Ef=0
\]
with general solution 
\[
F=F\bigg[(A_x+\eta')'-\frac{(A_x+\eta')}{\Ef}(\Ef)'\bigg]\,.
\]

The $N(N^x)'$ integral in (\ref{HDBracket1}) gives the additional condition
(assuming again that there are no derivatives of $\Kf$)
\begin{align}
-H_A&+\Par{H}{(A_x+\eta')}(A_x+\eta')+\CPar{H}{\Ef}\Ef \notag\\
+2\Par{H}{(A_x+\eta')'}&(A_x+\eta')'+2\CPar{H}{(\Ef)'}(\Ef)'+\CPar{H}{(E^x)'}(E^x)'=\mathcal{F}_3H  \label{HDBracketCond12}
\end{align}
an extension of (\ref{HDBracketCond00}).

On the other hand, the $\{H,H\}$ bracket (\ref{HHBracket}) is
\begin{align}
\{H[M],H[N]\}=2G\int \md x\,(MN'-NM')\bigg[&\gamma\bigg(\Delta_{E^x}^1-\CPar{H}{E^x\,'}\bigg)\bigg(\Par{H}{A_x}-\bigg(\Par{H}{A_x'}\bigg)'\bigg)\,+ \notag\\
&+\bigg(\Delta_{E^x}^0+\CPar{H}{E^x}-\bigg(\CPar{H}{(E^x)'}\bigg)'\bigg)\gamma\Par{H}{A_x'}\,+ \notag\\
&+\frac{1}{2}\bigg(\Delta_{\Ef}^1-\CPar{H}{(\Ef)'}\bigg)\bigg(\Par{H}{\Kf}-\bigg(\Par{H}{\Kf'}\bigg)'\bigg)\,+ \notag\\
&+\frac{1}{2}\bigg(\Delta^0_{\Ef}+\CPar{H}{\Ef}-\bigg(\CPar{H}{(\Ef)'}\bigg)'\bigg)\Par{H}{\Kf'}\bigg] \notag\\
+2G\int  \md x\,(MN''-NM'')&\bigg[\gamma\left(\Delta_{E^x}^2\right)\bigg(\Par{H}{A_x}-\bigg(\Par{H}{A_x'}\bigg)'\bigg)\notag\\
-2G\int \md x\,(M'N''-N'M''&)\gamma\Delta_{E^x}^2\Par{H}{A_x'}\,. \label{HHBracketFirstOrd}
\end{align}

The last integral, using (\ref{secondOrdCond2}), imposes the condition  
\begin{equation} \label{HHBracketCond1}
-2G\gamma\Delta_{E^x}^2\Par{H}{A_x'}=\mathcal{F}_4H+\mathcal{F}_5D\,,
\end{equation}
or, again explicitly
\begin{eqnarray}
&&-s\bar{\alpha}_\Gamma\frac{|E^x|^{\frac{1}{2}}}{\Ef}\bigg[
\frac{1}{\alpha(f_1+1)}\bigg(\Par{(\alpha(f_1+1))}{A_x'}\bigg)H_0
+\frac{1}{\bar{\alpha}f_2}\bigg(\Par{(\bar{\alpha}f_2)}{A_x'}\bigg)H_A 
\nonumber\\
&&\qquad\qquad 
+\frac{1}{\alpha_\Gamma}\bigg(\Par{\alpha_\Gamma}{A_x'}\bigg)H_\Gamma^1 
+\frac{1}{\bar{\alpha}_\Gamma}
\bigg(\Par{\bar{\alpha}_\Gamma}{A_x'}\bigg)H_\Gamma^{2,3}\bigg] 
=\mathcal{F}_4H+\mathcal{F}_5D. \label{HHBracketCond1e}
\end{eqnarray}
Once more, if we do not consider derivatives of $\Kf$, or we take the weaker
assumption $\partial \bar{\alpha}_\Gamma/\partial \Kf'=0$, noting that on the
left hand side only $H_\Gamma^{2}$ contains $(E^x)''$, we have
\begin{equation}
\mathcal{F}_4=-s\frac{|E^x|^\frac{1}{2}}{\Ef}\Par{\bar{\alpha}_\Gamma}{A_x'}\,.
\end{equation}
Hence, if we do not consider derivatives of $\Kf$ and if $\bar{\alpha}_\Gamma$
is independent of $A_x'$, we must have $\partial H/\partial A_x'=0$ even if we
allow for inverse-triad corrections. In particular this shows that radial
holonomy corrections alone cannot be anomaly-free at first order. According to
our calculations, the Hamiltonian constraint can depend on $A_x'$ only if it
also depends on $K_{\varphi}'$ or if correction functions depend on
$(E^{\varphi})'$. While we have not found such a consistent version (the next
section provides further insights), if one exists it would require tightly
related radial and angular holonomy corrections for all anomalies to cancel.

\section{Extrinsic-curvature expansion}
\label{s:ExtExp}

Some properties of consistent constraints can be derived by different,
somewhat more condensed methods if one makes use of more explicit expansions
of modified constraints. Using such methods, we now consider a derivative
order one above the classical one, allowing for derivatives of both $A_x$ (or
$K_x$) and $K_{\varphi}$.

In order to make spatial derivatives of extrinsic curvature in the functions
$f_1$ and $f_2$ of (\ref{ModifiedHamiltonian}) more explicit, we consider an
expansion of the form (\ref{Kderiv}) for these functions. We also assume time
reversibility, so that there are no terms with odd powers of the extrinsic
curvatures $K_x$ and $K_{\varphi}$. With these assumptions the Hamiltonian
density may be expanded as
\begin{equation} \label{HK}
H = \gul{00}{00} + \gul{11}{00}K_x K_{\varphi} + \gul{20}{00}K_x^{2} +\gul{02}{00}K_{\varphi}^{2} + \gul{10}{10}K_xK_x' + \gul{10}{01}K_x K_{\varphi}' + \gul{01}{01}K_{\varphi}K_{\varphi}' + \gul{01}{10}K_{\varphi}K_x'\,.
\end{equation}
with `coefficients' $\gul{ij}{kl}$ depending only on triad variables (and
their derivatives). For the classical constraint, we have
\begin{eqnarray}
\gul{00}{00} &=&
-\frac{1}{G}\left(\frac{E^{\varphi}}{\sqrt{|E^x|}}(1-\Gamma_{\varphi}^2)+ 
  2s\sqrt{|E^x|} \Gamma_{\varphi}'\right)\\
\gul{11}{00} &=& -\frac{1}{G}s\sqrt{|E^x|}\\
\gul{02}{00} &=& -\frac{1}{2G}\frac{E^{\varphi}}{\sqrt{|E^x|}}
\end{eqnarray}
and the rest zero.

As a short-cut, we will parameterize components and spatial derivatives of the
densitized triad by the sequence 
\begin{equation} \label{Esequence}
(E^x,E^{\varphi}, \Gamma_{\varphi},
\Gamma_{\varphi}',\Gamma_{\varphi}'',\ldots)
\end{equation}
where, we recall,
\begin{equation} \label{Gamma}
\Gamma_{\varphi}=-\frac{(E^x)'}{2E^{\varphi}}
\end{equation}
is a component of the classical spin connection. The sequence
(\ref{Esequence}) is not the most general set of derivatives of both $E^x$ and
$E^{\varphi}$. For our specific example, we assume that corrections appear as
powers or derivatives of $\Gamma_{\varphi}$, modeled on the form of
higher-curvature corrections. We do not attempt a more general analysis in
this paper. 
(If the hypersurface-deformation algebra is deformed, the classical
structure of Riemannian space-time does not apply and it is no longer clear
what a spin connection is that $\Gamma_{\varphi}$ could be related to by
pull-back. But the spatial structure remains unmodified in our setting, and in
any case we are still free to use the same $\Gamma_{\varphi}$ as a function of
the phase-space degrees of freedom $E^x$ and $E^{\varphi}$.)

We then have
\begin{eqnarray*}
\frac{\delta H(y)}{\delta E^x(z)} &=& \frac{\partial H(y)}{\partial
  E^x}\delta(y,z)
-\pard{H(y)}{\Gamma_{\varphi}}\frac{\delta'(y,z)}{2E^{\varphi}}
+ \pard{H(y)}{\Gamma_{\varphi}'}\left(-\frac{\delta''(y,z)}{2E^{\varphi}} +
\frac{(E^{\varphi})'\delta'(y,z)}{2(E^{\varphi})^{2}}\right) \\
&&+\pard{H(y)}{\Gamma_{\varphi}''}\left(-\frac{\delta'''(y,z)}{2E^{\varphi}} +
\frac{(E^{\varphi})'}{(E^{\varphi})^{2}}\delta''(y,z) +
\frac{(E^{\varphi})''}{2(E^{\varphi})^{2}}\delta'(y,z) -
\frac{((E^{\varphi})')^{2}}{(E^{\varphi})^{3}}\delta'(y,z)\right) \\
&=:& E_1(y)\delta(y,z)+ E_2(y)\delta'(y,z)+ E_3(y)\delta''(y,z)+ E_4(y)\delta'''(y,z)
\end{eqnarray*}
with
\begin{eqnarray}
 E_1 &=& \frac{\partial H}{\partial E^x}\,,\\
E_2 &=& -\frac{1}{2E^{\varphi}}\pard{H}{\Gamma_{\varphi}}+
\frac{(E^{\varphi})'}{2(E^{\varphi})^{2}}\pard{H}{\Gamma_{\varphi}'}   +
\left(\frac{(E^{\varphi})''}{2(E^{\varphi})^{2}} - 
\frac{((E^{\varphi})')^{2}}{(E^{\varphi})^{3}}\right)
\pard{H}{\Gamma_{\varphi}''} \,,\\
E_3&=& -\frac{1}{2E^{\varphi}}\pard{H}{\Gamma_{\varphi}'}+
\frac{(E^{\varphi})'}{(E^{\varphi})^{2}} \pard{H}{\Gamma_{\varphi}''}\,,\\
E_4 &=& -\frac{1}{2E^{\varphi}} \pard{H}{\Gamma_{\varphi}''}
\end{eqnarray}
and
\begin{eqnarray*}
\frac{\delta H(y)}{\delta E^{\varphi}(z)} &=& \frac{\partial H(y)}{\partial
  E^{\varphi}}\delta(y,z)+
\frac{(E^x)'}{2(E^{\varphi})^{2}}\pard{H(y)}{\Gamma_{\varphi}}\delta(y,z)\\
&&+ \pard{H(y)}{\Gamma_{\varphi}'}
\left(\frac{(E^{x})''}{2(E^{\varphi})^{2}}\delta(y,z) 
- \frac{(E^x)'(E^{\varphi})'}{(E^{\varphi})^{3}}\delta(y,z) +
\frac{(E^x)'}{2(E^{\varphi})^{2}}\delta'(y,z)\right) \\
&&+ \pard{H(y)}{\Gamma_{\varphi}''}
\left(\frac{(E^{x})'''}{2(E^{\varphi})^{2}}\delta(y,z) 
+ \frac{(E^{x})''}{(E^{\varphi})^{2}}\delta'(y,z) -
2\frac{(E^{x})''(E^{\varphi})'}{(E^{\varphi})^{3}}\delta(y,z) \right.\\
&&\qquad\qquad\;\left.   + \frac{(E^x)'}{2(E^{\varphi})^{2}}\delta''(y,z)
- \frac{(E^x)' (E^{\varphi})''}{(E^{\varphi})^{3}}\delta(y,z) + 3\frac{(E^x)'((E^{\varphi})')^{2}}{(E^{\varphi})^{4}}\delta(y,x) \right.\\
&&\qquad\qquad\;\left.  
 - 2\frac{(E^x)'(E^{\varphi})'}{(E^{\varphi})^{3}}\delta'(y,x)\right)\\
&=:& \bar{E}_1(y)\delta(y,z)+ \bar{E}_2(y)\delta'(y,z)+ \bar{E}_3(y)\delta''(y,z)
\end{eqnarray*}
with
\begin{eqnarray}
\bar{E}_1 &=& \frac{\partial H}{\partial E^{\varphi}}+
\frac{(E^x)'}{2(E^{\varphi})^{2}}\pard{H}{\Gamma_{\varphi}}
+\left(\frac{(E^{x})''}{2(E^{\varphi})^{2}} 
- \frac{(E^x)'(E^{\varphi})'}{(E^{\varphi})^{3}}\right) 
\pard{H}{\Gamma_{\varphi}'}\\
&&+\left(\frac{(E^{x})'''}{2(E^{\varphi})^{2}}-
  2\frac{(E^{x})''(E^{\varphi})'}{(E^{\varphi})^{3}}- \frac{(E^x)'
    (E^{\varphi})''}{(E^{\varphi})^{3}} +
  3\frac{(E^x)'((E^{\varphi})')^{2}}{(E^{\varphi})^{4}}\right) 
\pard{H}{\Gamma_{\varphi}''} \,, \nonumber\\
\bar{E}_2 &=& \frac{(E^x)'}{2(E^{\varphi})^{2}}\pard{H}{\Gamma_{\varphi}'}+
\left(\frac{(E^{x})''}{(E^{\varphi})^{2}}-
  2\frac{(E^x)'(E^{\varphi})'}{(E^{\varphi})^{3}}\right) 
\pard{H}{\Gamma_{\varphi}''}\,,\\
\bar{E}_3 &=&
\frac{(E^x)'}{2(E^{\varphi})^{2}} \pard{H}{\Gamma_{\varphi}''}\,.
\end{eqnarray}
Here and in what follows, derivatives of delta functions are taken with
respect to the first argument: $y$ for $\delta(y,z)$ above.

In the $\{H,H\}$ bracket, these functional derivatives will appear together
with those by $K_x$ and $K_{\varphi}$, respectively. The latter are
\begin{eqnarray*}
\frac{\delta H(x)}{\delta K_x(z)} &=& \left(\gul{11}{00}K_{\varphi}+
2\gul{20}{00}K_x+ \gul{10}{10}K_x'
 + \gul{10}{01}K_{\varphi}'\right)\delta(x,z)\\
&& + \left(\gul{10}{10}K_x+
 \gul{01}{10}K_{\varphi}\right)\delta'(x,z) \\
&=:& K_1(x)\delta(x,z)+K_2(x)\delta'(x,z)\\
\frac{\delta H(x)}{\delta K_{\varphi}(z)} &=& \left(\gul{11}{00}K_x +
2\gul{02}{00}K_{\varphi}
 + \gul{01}{01}K_{\varphi}'+
 \gul{01}{10}K_x'\right)\delta(x,z) \\
&& + \left(\gul{10}{01}K_x + \gul{01}{01}K_{\varphi}\right)\delta'(x,z)\\
&:=& \bar{K}_1(x)\delta(x,z)+\bar{K}_2(x)\delta'(x,z) \,.
\end{eqnarray*}

With these preparations, integrating out the $x$ and $y$ dependence with
smearing functions $M(x)$ and $N(y)$, we write
\begin{align}
\frac{1}{G}\{H[M]&,H[N]\} = \notag\\
=&\int\int\int\md x\,\md y\, \md z\,M(x)N(y)
\bigg(2\fund{H(x)}{K_x(z)}\fund{H(y)}{E^x(z)}+\fund{H(x)}{\Kf(z)}\fund{H(y)}{\Ef(z)} \, - \, (x \leftrightarrow y)\bigg)\notag\\ 
=& \int\int\int\md x\,\md y\, \md z\,M(x)N(y)
\bigg(2\fund{H(x)}{K_x(z)}\fund{H(y)}{E^x(z)} +
\fund{H(x)}{K_{\varphi}(z)}\fund{H(y)}{E^{\varphi}(z)}\bigg) - (M \leftrightarrow N) \notag\\
=&\int\md z\,\big[2\big(MNK_1E_1-M(NE_2)'K_1+M(NE_3)''K_1-M(NE_4)'''K_1  \notag\\
&\qquad\quad -(MK_2)'E_1+(MK_2)'(NE_2)'-(MK_2)'(NE_3)''+(MK_2)'(NE_4)'''\big)\notag\\
&\qquad\; +MN\bar{K}_1\bar{E}_1-M(N\bar{E}_2)'\bar{K}_1+M(N\bar{E}_3)''\bar{K}_1  \notag\\
&\qquad\; -(M\bar{K}_2)'\bar{E}_1+(M\bar{K}_2)'(N\bar{E}_2)'-(M\bar{K}_2)'(N\bar{E}_3)''\big]\quad- (M \leftrightarrow N) \notag\\
=&\int\md z\,(MN'-NM')\big[\,2\big(-K_1E_2+K_2E_1+2K_1E_3'-K_2E_2'+K_2'E_2-3K_1E_4'' \notag\\
&\qquad\qquad\qquad\qquad\quad+K_2E_3''-2K_2'E_3'+3K_2'E_4''-K_2E_4'''\big)  
-\bar{K}_1\bar{E}_2+\bar{K}_2\bar{E}_1  \notag\\
&\qquad\qquad\qquad\qquad\quad+2\bar{K}_1\bar{E}_3'-\bar{K}_2\bar{E}_2'+\bar{K}_2'\bar{E}_2 
+\bar{K}_2\bar{E}_3''-2\bar{K}_2'\bar{E}_3' \,\big] \notag\\
&+\int\md z\,(MN''-NM'')\big[\,2\big(K_1E_3-3K_1E_4'-K_2'E_3+3K_2'E_4'\big)+\bar{K}_1\bar{E}_3-\bar{K}_2'\bar{E}_3\,\big] \notag\\
&+\int\md z\,(M'N''-N'M'')\big[\,2\big(-K_2E_3+3K_2E_4'\big)-\bar{K}_2\bar{E}_3\,\big]  \notag\\
&+\int\md z\,(MN'''-NM''')\big(K_2'-K_1\big)E_4 \notag\\
&+\int\md z\,(M'N'''-N'M''')K_2E_4
\end{align}

We are now ready to apply (\ref{hthree2}) and (\ref{hthree3}) to read off two
independent conditions. The second one of these conditions provides a lengthy
equation, but coefficients of $\Kf'''$ and $K_x'''$ on the left-hand side of
(\ref{hthree3}) must vanish since these cannot be matched with a linear
combination of constraints. The coefficient of $K_x'''$ is identically zero,
but the requirement of vanishing $\Kf'''$-term gives the equation
\begin{equation}  \label{Kx3condition}
E_4(\gul{01}{10}-\gul{10}{01})=0.
\end{equation} 

We first consider the simplest possibility $E_4=0$. With $E_4=0$, we have
$\partial H/\partial\Gamma_{\varphi}''=0$ and therefore
$\bar{E}_3=0$. Condition (\ref{hthree2}) then has only one non-zero term
$K_2E_3\approx0$, which implies $K_2=0$. (There is no $K_{\varphi}$ in the
diffeomorphism constraint but only its first derivative, and $E_3$ cannot be
zero because it does not vanish classically.) With $K_2=0$, we must have
$\gul{10}{10}=0=\gul{01}{10}$, and there cannot be first-order derivatives of
$K_x$ in the Hamiltonian.  A consistent version with radial holonomies seems
to require higher than next order in derivatives of the triad.

It remains to evaluate all remaining terms of (\ref{hthree3}):
\begin{equation}
2K_1E_2+\bar{K}_1\bar{E}_2-\bar{K}_2\bar{E}_1+2K_1'E_3-2K_1E_3'
+\bar{K}_2\bar{E}_2'-\bar{K}_2'\bar{E}_2\approx 0\,.
\end{equation}
For second derivatives of $K_x$ and $K_{\varphi}$ to be absent (from $K_1'$),
we must have $\gul{10}{01}=0$ (in addition to $\gul{10}{10}$ which we have
already derived).  We write out the remaining $K$-terms explicitly:
\begin{align*}
&\left(2\gul{11}{00}(E_2-E_3')+2\gul{02}{00}\bar{E}_2-\gul{01}{01}(\bar{E}_1-\bar{E}_2')
+2(\gul{11}{00})'E_3-(\gul{01}{01})'\bar{E}_2\right)\Kf  \\
&+2\gul{11}{11}\bar{E}_3\Kf'+\left(4\gul{20}{00}(E_2-E_3')+\gul{11}{00}\bar{E}_2+4(\gul{20}{00})'E_3\right)K_x+
4\gul{20}{00}\bar{E}_3K_x' \approx 0\,. 
\end{align*}
Since there are no terms independent of the $K$ components, this expression
(assuming $\gul{00}{00}\neq 0$) must be proportional to the diffeomorphism
constraint only: $\mathcal{F}\left(2E^{\varphi}K_{\varphi}'-K_x(E^x)'\right)$.
For $K_x'$ to be absent, we must have $\gul{20}{00}=0$, and comparison of the
$K_x$- and $\Kf'$-terms gives the same consistent relation
\begin{equation}
\mathcal{F}=-\frac{1}{2(\Ef)^{2}}\Par{H}{\Gamma_{\varphi}'}\gul{11}{00}\,.
\end{equation}
This function plays the role of the deformation function in the
constraint algebra. It might differ from the previous form of $\beta$
if the correction functions depend non-trivially on
$\Gamma_{\varphi}'$ in the presence of derivative corrections.

The $K_{\varphi}$-coefficient must vanish, which provides one further
consistency condition
\begin{equation} \label{Cond}
2\gul{11}{00}(E_2-E_3')+2\gul{02}{00}\bar{E}_2
-\gul{01}{01}(\bar{E}_1-\bar{E}_2') 
+2(\gul{11}{00})'E_3-(\gul{01}{01})'\bar{E}_2=0\,.
\end{equation}
This equation can be interpreted as a condition for $\gul{01}{01}$ for
given $\gul{11}{00}$ and $\gul{02}{00}$,  and the latter two
functions remain free. (Seen in this way, (\ref{Cond}) is a
first-order ordinary differential equation for $\gul{01}{01}(x)$. The
integration constant is fixed by the boundary condition that the
classical limit $\gul{01}{01}=0$ should be reached at spatial
infinity.) It is then possible to have a derivative correction
proportional to $K_{\varphi}K_{\varphi}'$, but the coefficient is
strictly related to the correction functions multiplying the classical
curvature terms. The derivative term affects the deformation function
$\beta$ only indirectly via $\gul{11}{00}$ and (\ref{Cond}).

Instead of solving for $\gul{01}{01}$, we may rewrite (\ref{Cond}) as
\begin{equation} \label{Cond2}
\gul{02}{00}= \frac{1}{2}(\gul{01}{01})'
-\frac{1}{2G} \frac{E^{\varphi}}{\sqrt{|E^x|}}  \alpha
\end{equation}
with
\begin{equation}
 \alpha = -\frac{G\sqrt{|E^x|}}{E^{\varphi}}
\left(\gul{01}{01}\frac{\bar{E}_1-\bar{E}_2'}{\bar{E}_2}- 
2\gul{11}{00}\frac{E_2-E_3'}{\bar{E}_2}- 
2(\gul{11}{00})'\frac{E_3}{\bar{E}_2}\right)\,.
\end{equation}
Inserting (\ref{Cond2}) in (\ref{HK}) shows that $\alpha$ plays the
same role as in (\ref{ModifiedHamiltonian}) and, more importantly,
that the new derivative term in the Hamiltonian density $H$ can only
be a total derivative $\frac{1}{2}(\gul{01}{01} K_{\varphi}^2)'$. The
allowed derivative correction is therefore of a very special form that
is not necessarily expected from general holonomy
corrections. Nevertheless, the correction would non-trivially affect
the solution space of the theory because the Hamiltonian density, upon
multiplication with a general lapse function, is not just affected by
a total derivative.

As a consistency check, we show that the previously known consistent versions
satisfy (\ref{Cond}). If there are no derivative corrections, we have the 
parameterization (\ref{ModifiedHamiltonian}) with coefficients independent of
$\Gamma_{\varphi}$ and its derivatives:
\begin{eqnarray}
 \gul{11}{00} = -\frac{1}{G} \sqrt{|E^x|}\bar{\alpha}\quad&,&\quad
\gul{02}{00} = -\frac{1}{2G} \frac{E^{\varphi}}{\sqrt{|E^x|}}\alpha\,,\\
 \frac{\partial H}{\partial\Gamma_{\varphi}} = -\frac{1}{2G}
 \frac{(E^x)'}{\sqrt{|E^x|}} \alpha_{\Gamma}\quad&,&\quad
\frac{\partial H}{\partial\Gamma_{\varphi}'} = -\frac{1}{G} \sqrt{|E^x|}
\bar{\alpha}_{\Gamma}\,.   \label{NogammaDependence}
\end{eqnarray}
With the equations found here, we obtain the deformation function
\begin{equation}
 \beta=\bar{\alpha}\bar{\alpha}_{\Gamma}
\end{equation}
and one consistency condition
\begin{equation}
 \alpha\bar{\alpha}_{\Gamma}- \bar{\alpha}\alpha_{\Gamma}-\frac{2|E^x|}{(E^x)'}\left(\bar{\alpha}'\bar{\alpha}_{\Gamma}- \bar{\alpha}\bar{\alpha}_{\Gamma}'\right)=0\,.
\end{equation}
This equation reproduces the condition found in \cite{JR}.

Finally, following similar arguments, it is not difficult to see that
the other possibility to fulfill condition (\ref{Kx3condition}):
$\gul{01}{10}=\gul{10}{01}$ is not consistent. Therefore, $E_4=0$ and
we cannot have any dependence on $\Gamma_{\varphi}''$ to this order.
Since $\Gamma_{\varphi}''$ depends on $(E^{\varphi})''$, the
result that $\partial H/\partial \Gamma_{\varphi}''=0$ is related to
the condition found in Sec.~\ref{s:Diffeo}, but here we are testing for
a combination of $(E^{\varphi})''$ and $(E^x)'''$.

\section{Conclusions}

The anomaly problem is one of the most crucial issues in canonical and loop
quantum gravity. If it cannot be resolved, canonical quantum gravity cannot be
shown to be consistent. Unfortunately, the problem is also one of the most
complicated ones, and therefore any result in this direction is useful. In
some models, one can make progress with commutators of operators
\cite{ThreeDeform,TwoPlusOneDef,TwoPlusOneDef2,AnoFreeWeak}, but except for
the $2+1$-dimensional example in \cite{ThreeDeform} it remains difficult to
tackle non-local holonomies. Moreover, the step from operator equations to
geometrical statements is non-trivial, requiring some handle on semiclassical
states which constitutes another difficult and important problem of canonical
quantum gravity. Effective methods, as further developed in this article,
sidestep many of these complications and can still provide fundamental
insights.

If canonical quantum gravity gives rise to a consistent operator version of
the constraint algebra, expectation values lead to a consistent effective
version of constraints. By ruling out certain terms in effective constraints,
one can therefore conclude that the possible form of consistent operators is
restricted. We have done just that in the present paper, by limiting (but
  not completely ruling out) the connection dependence to next-to-leading
order in an expansion by spatial derivatives.

We have not attempted to go beyond the next-to-leading order, given the
complexity, but our methods are suitable for such a task. We do not speculate
on the verdict whether a derivative expansion of holonomies can be implemented
consistently. But it is of interest to spell out what implications a negative
result would have. (The implications of a positive verdict are obvious.) If
higher spatial derivatives can be ruled out, loop quantum gravity as it is
commonly understood would be shown to be inconsistent. However, a more careful
view could still be possible, a view that relies on a combination of higher
spatial with higher time derivatives. Even though this consequence would agree
with the expectation from higher-curvature effective actions, it is not
necessarily implied by loop quantum gravity, as we have discussed in detail in
Sec.~\ref{s:Deriv}. Holonomy modifications of the Hamiltonian constraint by
themselves do not imply a close link with higher time derivatives. If such a
link would be required by the condition of anomaly-freedom, it would mean that
the construction of Hamiltonian constraint operators must be tightly connected
to the form of the allowed dynamical states. After all, higher time
derivatives arise in canonical quantum theories from quantum back-reaction of
moments of a state. If holonomy modifications can be made consistent only with
a careful choice of corrections that contain higher time derivatives, the form
of states on which the constraint algebra can be represented must be
non-trivially restricted; not all kinematical states could be allowed. One
would have to find a domain of states smaller than the kinematical Hilbert
space but larger than the physical one (since one aims to represent the
off-shell algebra). Such classes of states have not been considered yet in
loop quantum gravity, but, given the difficulties in finding consistent
effective realizations with non-pointwise holonomies, they might be the only
way to make the theory consistent.

Another, perhaps more dramatic consequence would apply to loop quantum
cosmology. If consistent holonomy corrections require closely related higher
time derivatives, the current cosmological models used in this context are
wrong: The usual modified Friedmann equations do not contain higher time
derivatives even though such terms would be of a similar magnitude as the
modifications.  Results obtained by solving for full wave functions implicitly
contain higher time derivatives via quantum back-reaction. However, in a
mini\-superspace model these implicit corrections are not tied to holonomy
modifications in a strong-enough way to ensure anomaly-freedom.  While higher
spatial derivatives do not matter for minisuperspace equations, higher time
derivatives are important terms that could change the implications claimed for
holonomy modifications in models in which a consistent embedding in
inhomogeneous geometries has not been confirmed. In the cosmological context,
it is also important to mention the mounting evidence for signature change at
high curvature. Our equation (\ref{betaC1}) confirms and strengthens the
phenomenon in spherically symmetric models, in accordance with a similar
result recently obtained for cosmological perturbations \cite{ScalarHolInv}.

We repeat that our results do not suffice to rule out consistent holonomy
corrections. Our last remarks are meant to show the importance of the anomaly
problem, a question which is ignored in most of the ``physical'' results
claimed by the theory in cosmological or black-hole models. 

\section*{Acknowledgements}

This work was supported in part by NSF grants PHY-0748336, PHY-1307408 and
NSF-CONACyT Grant: Strong Back Reaction Effects in Quantum Cosmology.

\begin{appendix}

\section{Conditions from antisymmetric higher-derivative multipliers}

We show here how  to rewrite the general expression (\ref{antisymHHbracket})
in the form $\sum_{j=0}^{2n-1}\int \md x\, NM^{(j)}g_j$.

First, integrating by parts and using (\ref{binomialIdentity}) and (\ref{PascalSum}),
we  have 
\[
\sum_{j=1}^n\sum_{i=0}^{j-1}\int \md x\,M^{(i)}N^{(j)}h_{i,j}=\sum_{j=1}^n\sum_{k=0}^{j-1}\int \md x\,MN^{(j+k)}\sum_{l=0}^{j-k-1}(-1)^{k+l}\binom{k+l}{l}h_{k+l,j}^{(l)}\,,
\]
and
\[
-\sum_{j=1}^n\sum_{i=0}^{j-1}\int \md x\,M^{(j)}N^{(i)}h_{i,j}=\sum_{j=1}^n\sum_{k=0}^{j-1}\sum_{l=0}^j\int \md x\,MN^{(j+k-l)}(-1)^{j+1}\binom{j}{l}h_{k,j}^{(l)}\,.
\]
Adding these two expressions results in
\begin{align}
\sum_{j=1}^n\sum_{i=0}^{j-1}\int \md x\,&(M^{(i)}N^{(j)}-N^{(i)}M^{(j)})h_{i,j}= \notag\\
=&\sum_{j=1}^n\sum_{k=0}^{j-1}\int \md x\,MN^{(j+k)}\Bigg((-1)^{j+1}h_{k,j}+\sum_{l=0}^{j-k-1}(-1)^{k+l}\binom{k+l}{l}h_{k+l,j}^{(l)}\Bigg) \notag\\
&+\sum_{j=1}^n\sum_{k=0}^{j-1}\sum_{l=1}^j\int \md x\,MN^{(j+k-l)}(-1)^{j+1}\binom{j}{l}h_{k,j}^{(l)}\,.
\end{align}
Defining $s:=j+k$, the first line of the right hand side of this expression may be written as
\begin{align}
&\sum_{j=1}^n\sum_{k=0}^{j-1}\int \md x\,MN^{(j+k)}\Bigg((-1)^{j+1}h_{k,j}+\sum_{l=0}^{j-k-1}(-1)^{k+l}\binom{k+l}{l}h_{k+l,j}^{(l)}\Bigg)= \notag\\
&=\sum_{s=1}^{2n-1}\int \md x\,MN^{(s)}\sum_{j=\,\li s/2 \ri+1}^{\text{min}(s,n)}\left((-1)^{j+1}h_{s-j,j}+\sum_{l=0}^{2j-s-1}(-1)^{s-j+l}\binom{s-j+l}{l}h_{s-j+l,j}^{(l)}\right)\,,
\end{align}
where $\li s/2 \ri$ denotes the integer part of $s/2$. The second line requires a little more work: defining $r:=k-l+1$ and then $s:=j+r-1$, we have
\begin{align}
&\sum_{j=1}^n\sum_{k=0}^{j-1}\sum_{l=1}^j\int \md x\,MN^{(j+k-l)}(-1)^{j+1}\binom{j}{l}h_{k,j}^{(l)}= \notag\\
&=\sum_{j=1}^n\sum_{r=-(j-1)}^{j-1}\int \md x\, MN^{(j+r-1)}(-1)^{j+1}\sum_{k=\text{max}(r,0)}^{\text{min}(j+r-1,j-1)}\binom{j}{k-r+1}h_{k,j}^{(k-r+1)} \notag\\
&=\sum_{s=0}^{2n-2}\int \md x\,MN^{(s)}\sum_{j=\,\li\frac{s+1}{2}\ri+1}^n(-1)^{j+1}\sum_{k=\text{max}(s-j+1,0)}^{\text{min}(s,j-1)}\binom{j}{k+j-s}h_{k,j}^{(k+j-s)}\,.
\end{align}
Combining these results we finally get
\begin{align}
\sum_{j=1}^n\sum_{i=0}^{j-1}&\int \md x\,(M^{(i)}N^{(j)}-N^{(i)}M^{(j)})h_{i,j}= \notag\\
&=\sum_{s=1}^{2n-1}\int \md x\,MN^{(s)}\sum_{j=\,\li s/2
  \ri+1}^{\text{min}(s,n)}\left((-1)^{j+1}h_{s-j,j}+\sum_{l=0}^{2j-s-1}(-1)^{s-j+l}\binom{s-j+l}{l}h_{s-j+l,j}^{(l)}\right)+
\notag\\ 
&+\sum_{s=0}^{2n-2}\int \md
x\,MN^{(s)}\sum_{j=\,\li\frac{s+1}{2}\ri+1}^n(-1)^{j+1}\sum_{l=\text{max}(s-j+1,0)}^{\text{min}(s,j-1)}\binom{j}{l+j-s}h_{l,j}^{(l+j-s)}\,. 
\end{align}

This gives $2n$ equations from which there will be  $n$ independent
conditions. We have (for $n\ge 2$): 
\[
g_0=\sum_{j=1}^n(-1)^{j+1}h_{0,j}^{(j)}\approx0\,,
\]
\begin{align}
g_s=\sum_{j=\,\li s/2 \ri+1}^{\text{min}(s,n)}\left((-1)^{j+1}h_{s-j,j}+\sum_{l=0}^{2j-s-1}(-1)^{s-j+l}\binom{s-j+l}{l}h_{s-j+l,j}^{(l)}\right)+ \notag\\
+\sum_{j=\,\li\frac{s+1}{2}\ri+1}^n(-1)^{j+1}\sum_{l=\text{max}(s-j+1,0)}^{\text{min}(s,j-1)}\binom{j}{l+j-s}h_{l,j}^{(l+j-s)}\approx0\,, \notag
\end{align}
for $s=1,\dots,2n-2$, and
\[
g_{2n-1}=(-1)^{n-1}2h_{n-1,n}\approx0\,.
\]

\end{appendix}


\begin{thebibliography}{10}

\bibitem{UniformDisc}
M.\ Campiglia, C.\ Di~Bartolo, R.\ Gambini, and J.\ Pullin,
\newblock Uniform discretizations: a new approach for the quantization of
  totally constrained systems,
\newblock {\em Phys.\ Rev.\ D} 74 (2006) 124012, [gr-qc/0610023]

\bibitem{PerfectAction}
B.\ Bahr and B.\ Dittrich,
\newblock Improved and Perfect Actions in Discrete Gravity,
\newblock {\em Phys.\ Rev.\ D} 80 (2009) 124030, [arXiv:0907.4323]

\bibitem{BrokenAction}
B.\ Bahr and B.\ Dittrich,
\newblock Breaking and restoring of diffeomorphism symmetry in discrete gravity
\newblock (2009), [arXiv:0909.5688]

\bibitem{VectorHol}
J.\ Mielczarek, A.\ Cailleteau, Barrau, T.\, and J.\ Grain,
\newblock Anomaly-free vector perturbations with holonomy corrections in loop
  quantum cosmology,
\newblock {\em Class.\ Quant.\ Grav.} 29 (2012) 085009, [arXiv:1106.3744]

\bibitem{ScalarHol}
T.\ Cailleteau, J.\ Mielczarek, A.\ Barrau, and J.\ Grain,
\newblock Anomaly-free scalar perturbations with holonomy corrections in loop
  quantum cosmology,
\newblock {\em Class.\ Quant.\ Grav.} 29 (2012) 095010, [arXiv:1111.3535]

\bibitem{VectorSecond}
J.-P.\ Wu and Y.\ Ma,
\newblock Anomaly freedom of the vector modes with holonomy corrections in
  perturbative Euclidean loop quantum gravity,
\newblock {\em Phys.\ Rev.\ D} 86 (2012) 124044, [arXiv:1209.2766]

\bibitem{Rov}
C.\ Rovelli,
\newblock {\em Quantum Gravity},
\newblock Cambridge University Press, Cambridge, UK, 2004

\bibitem{ThomasRev}
T.\ Thiemann,
\newblock {\em Introduction to Modern Canonical Quantum General Relativity},
\newblock Cambridge University Press, Cambridge, UK, 2007, [gr-qc/0110034]

\bibitem{ALRev}
A.\ Ashtekar and J.\ Lewandowski,
\newblock Background independent quantum gravity: A status report,
\newblock {\em Class.\ Quantum Grav.} 21 (2004) R53--R152, [gr-qc/0404018]

\bibitem{Energy}
M.\ Bojowald, G.\ Hossain, M.\ Kagan, and C.\ Tomlin,
\newblock Quantum matter in quantum space-time,
\newblock {\em Quantum Matter} 2 (2013) 436--443, [arXiv:1302.5695]

\bibitem{Action}
M.\ Bojowald and G.~M.\ Paily,
\newblock Deformed General Relativity and Effective Actions from Loop Quantum
  Gravity,
\newblock {\em Phys.\ Rev.\ D} 86 (2012) 104018, [arXiv:1112.1899]

\bibitem{ThreeDeform}
A.\ Perez and D.\ Pranzetti,
\newblock On the regularization of the constraints algebra of Quantum Gravity
  in $2+1$ dimensions with non-vanishing cosmological constant,
\newblock {\em Class.\ Quantum Grav.} 27 (2010) 145009, [arXiv:1001.3292]

\bibitem{JR}
J.~D.\ Reyes,
\newblock {\em Spherically Symmetric Loop Quantum Gravity: Connections to
  2-Dimensional Models and Applications to Gravitational Collapse},
\newblock PhD thesis, The Pennsylvania State University, 2009

\bibitem{LTBII}
M.\ Bojowald, J.~D.\ Reyes, and R.\ Tibrewala,
\newblock Non-marginal LTB-like models with inverse triad corrections from loop
  quantum gravity,
\newblock {\em Phys.\ Rev.\ D} 80 (2009) 084002, [arXiv:0906.4767]

\bibitem{ConstraintAlgebra}
M.\ Bojowald, G.\ Hossain, M.\ Kagan, and S.\ Shankaranarayanan,
\newblock Anomaly freedom in perturbative loop quantum gravity,
\newblock {\em Phys.\ Rev.\ D} 78 (2008) 063547, [arXiv:0806.3929]

\bibitem{ScalarGaugeInv}
M.\ Bojowald, G.\ Hossain, M.\ Kagan, and S.\ Shankaranarayanan,
\newblock Gauge invariant cosmological perturbation equations with corrections
  from loop quantum gravity,
\newblock {\em Phys.\ Rev.\ D} 79 (2009) 043505, [arXiv:0811.1572]

\bibitem{AreaVol}
C.\ Rovelli and L.\ Smolin,
\newblock Discreteness of Area and Volume in Quantum Gravity,
\newblock {\em Nucl.\ Phys.\ B} 442 (1995) 593--619, [gr-qc/9411005],
\newblock Erratum: {\em Nucl.\ Phys.\ B} 456 (1995) 753

\bibitem{RS:Ham}
C.\ Rovelli and L.\ Smolin,
\newblock The physical Hamiltonian in nonperturbative quantum gravity,
\newblock {\em Phys.\ Rev.\ Lett.} 72 (1994) 446--449, [gr-qc/9308002]

\bibitem{QSDI}
T.\ Thiemann,
\newblock Quantum Spin Dynamics {(QSD)},
\newblock {\em Class.\ Quantum Grav.} 15 (1998) 839--873, [gr-qc/9606089]

\bibitem{RS:Spinnet}
C.\ Rovelli and L.\ Smolin,
\newblock Spin Networks and Quantum Gravity,
\newblock {\em Phys.\ Rev.\ D} 52 (1995) 5743--5759

\bibitem{AnoFree}
T.\ Thiemann,
\newblock Anomaly-Free Formulation of Non-Perturbative,
  Four-Dimensional Lorentzian Quantum Gravity,
\newblock {\em Phys.\ Lett.\ B} 380 (1996) 257--264, [gr-qc/9606088]

\bibitem{TwoPlusOneDef}
A.\ Henderson, A.\ Laddha, and C.\ Tomlin,
\newblock Constraint algebra in LQG reloaded : Toy model of a ${\rm U}(1)^{3}$
  Gauge Theory I,
\newblock {\em Phys.\ Rev.\ D} 88 (2013) 044028, [arXiv:1204.0211]

\bibitem{TwoPlusOneDef2}
A.\ Henderson, A.\ Laddha, and C.\ Tomlin,
\newblock Constraint algebra in LQG reloaded : Toy model of an Abelian gauge
  theory - II Spatial Diffeomorphisms,
\newblock {\em Phys.\ Rev.\ D} 88 (2013) 044029, [arXiv:1210.3960]

\bibitem{AnoFreeWeak}
C.\ Tomlin and M.\ Varadarajan,
\newblock Towards an Anomaly-Free Quantum Dynamics for a Weak Coupling Limit of
  Euclidean Gravity,
\newblock {\em Phys.\ Rev.\ D} 87 (2013) 044039, [arXiv:1210.6869]

\bibitem{SymmRed}
M.\ Bojowald and H.~A.\ Kastrup,
\newblock Symmetry Reduction for Quantized Diffeomorphism Invariant Theories of
  Connections,
\newblock {\em Class.\ Quantum Grav.} 17 (2000) 3009--3043, [hep-th/9907042]

\bibitem{SphSymm}
M.\ Bojowald,
\newblock Spherically Symmetric Quantum Geometry: States and Basic Operators,
\newblock {\em Class.\ Quantum Grav.} 21 (2004) 3733--3753, [gr-qc/0407017]

\bibitem{SphSymmHam}
M.\ Bojowald and R.\ Swiderski,
\newblock Spherically Symmetric Quantum Geometry: Hamiltonian Constraint,
\newblock {\em Class.\ Quantum Grav.} 23 (2006) 2129--2154, [gr-qc/0511108]

\bibitem{Springer}
M.\ Bojowald,
\newblock {\em Quantum Cosmology: A Fundamental Theory of the Universe},
\newblock Springer, New York, 2011

\bibitem{APSII}
A.\ Ashtekar, T.\ Pawlowski, and P.\ Singh,
\newblock Quantum Nature of the Big Bang: Improved dynamics,
\newblock {\em Phys.\ Rev.\ D} 74 (2006) 084003, [gr-qc/0607039]

\bibitem{InhomLattice}
M.\ Bojowald,
\newblock Loop quantum cosmology and inhomogeneities,
\newblock {\em Gen.\ Rel.\ Grav.} 38 (2006) 1771--1795, [gr-qc/0609034]

\bibitem{CosConst}
M.\ Bojowald,
\newblock The dark side of a patchwork universe,
\newblock {\em Gen.\ Rel.\ Grav.} 40 (2008) 639--660, [arXiv:0705.4398]

\bibitem{cosmoIV}
M.\ Bojowald,
\newblock Loop Quantum Cosmology IV: Discrete Time Evolution,
\newblock {\em Class.\ Quantum Grav.} 18 (2001) 1071--1088, [gr-qc/0008053]

\bibitem{SemiClass}
M.\ Bojowald,
\newblock The Semiclassical Limit of Loop Quantum Cosmology,
\newblock {\em Class.\ Quantum Grav.} 18 (2001) L109--L116, [gr-qc/0105113]

\bibitem{Simon}
J.~Z.\ Simon,
\newblock Higher-derivative Lagrangians, nonlocality, problems, and solutions,
\newblock {\em Phys.\ Rev.\ D} 41 (1990) 3720--3733

\bibitem{EffAc}
M.\ Bojowald and A.\ Skirzewski,
\newblock Effective Equations of Motion for Quantum Systems,
\newblock {\em Rev.\ Math.\ Phys.} 18 (2006) 713--745, [math-ph/0511043]

\bibitem{Karpacz}
M.\ Bojowald and A.\ Skirzewski,
\newblock Quantum Gravity and Higher Curvature Actions,
\newblock {\em Int.\ J.\ Geom.\ Meth.\ Mod.\ Phys.} 4 (2007) 25--52,
  [hep-th/0606232],
\newblock Proceedings of ``Current Mathematical Topics in Gravitation and
  Cosmology'' (42nd Karpacz Winter School of Theoretical Physics), Ed.\
  Borowiec, A.\ and Francaviglia, M.

\bibitem{HigherTime}
M.\ Bojowald, S.\ Brahma, and E.\ Nelson,
\newblock Higher time derivatives in effective equations of canonical quantum
  systems,
\newblock {\em Phys.\ Rev.\ D} 86 (2012) 105004, [arXiv:1208.1242]

\bibitem{HigherCurvHam}
N.\ Deruelle, M.\ Sasaki, Y.\ Sendouda, and D.\ Yamauchi,
\newblock Hamiltonian formulation of $f({\rm Riemann})$ theories of gravity,
\newblock {\em Prog.\ Theor.\ Phys.} 123 (2009) 169--185, [arXiv:0908.0679]

\bibitem{NonAb}
M.\ Bojowald,
\newblock Mathematical structure of loop quantum cosmology: Homogeneous models,
\newblock {\em SIGMA} 9 (2013) 082, [arXiv:1206.6088]

\bibitem{SchwarzN}
M.\ Bojowald, D.\ Cartin, and G.\ Khanna,
\newblock Lattice refining loop quantum cosmology, anisotropic models and
  stability,
\newblock {\em Phys.\ Rev.\ D} 76 (2007) 064018, [arXiv:0704.1137]

\bibitem{ScalarHolInv}
T.\ Cailleteau, L.\ Linsefors, and A.\ Barrau,
\newblock Anomaly-free perturbations with inverse-volume and holonomy
  corrections in Loop Quantum Cosmology, [arXiv:1307.5238]

\end{thebibliography}

\end{document}